\documentclass[english,10pt,final,twocolumn]{elsarticle}
\usepackage[T1]{fontenc}
\usepackage[latin9]{inputenc}
\usepackage{pdfpages}
\usepackage{amstext}
\usepackage{graphicx}
\usepackage{gensymb}
\usepackage{amssymb}
\usepackage{placeins}
\usepackage{subfigure}
\usepackage[title]{appendix}
\makeatletter
\usepackage{float}

\usepackage{blindtext}
\usepackage{titlesec}

\titleformat{\section}
  {\bfseries\large}{\thesection}{1em}{}
\titleformat{\subsection}
  {\bfseries}{\thesubsection}{1em}{}
  \titleformat{\appendix}
  {\bfseries\large}{\thesection}{1em}{}

\newcommand{\lyxmathsym}[1]{\ifmmode\begingroup\def\b@ld{bold}
  \text{\ifx\math@version\b@ld\bfseries\fi#1}\endgroup\else#1\fi}
\usepackage{braket}
\usepackage{mathrsfs}
\usepackage{slashed}
\usepackage{bm}
\usepackage{datetime}
\usepackage{siunitx}

\usepackage{amsmath}

\DeclareSIUnit[number-unit-product = {}]\clight{c}
\DeclareSIUnit\eVperc{\eV\per\clight}
\DeclareSIUnit\GeVpercs{\giga\eV\squared\per\clight\squared}
\DeclareSIUnit\MeVpercs{\mega\eV\per\clight\squared}
\journal{Physics Letters B}
\biboptions{numbers,sort&compress}
\usepackage[left=1.6cm,right=1.04cm,top=1.65cm,bottom=1.65cm,columnsep=25pt]{geometry}
\usepackage{lineno}
\usepackage{hyperref}  
\hypersetup{breaklinks = true, colorlinks = true, allcolors = magenta}
\usepackage{setspace}
\usepackage{graphicx}
\usepackage{wrapfig}

\makeatother

\usepackage{babel}
\begin{document}

\begin{frontmatter}{}

\title{
Measurements of the induced 
polarization in the quasi-elastic $A(e,e'\vec p\,)$ process in non-coplanar
kinematics.
}

\author[TAU]{S.J.~Paul\corref{cor2}\fnref{ucr}}
\ead{paulsebouh@mail.tau.ac.il}
\author[JSI]{T.~Kolar}
\author[JSI]{T.~Brecelj}

\author[Mainz]{P.~Achenbach}
\author[Mainz]{H.~Arenh\"ovel}
\author[TAU]{A.~Ashkenazi}
\author[JSI]{J.~Beri\v{c}i\v{c}}
\author[Mainz]{R.~B\"ohm}
\author[zagreb]{D.~Bosnar}
\author[Rutgers]{E.~Cline}
\author[TAU]{E.O.~Cohen}
\author[JSI]{L.~Debenjak}
\author[Mainz]{M.O.~Distler}
\author[Mainz]{A.~Esser}
\author[zagreb]{I.~Fri\v{s}\v{c}i\'{c}\fnref{mit}}
\author[Rutgers]{R.~Gilman}
\author[Pavia]{C.~Giusti}
\author[Konstanz]{M.~Heilig}
\author[Mainz]{M.~Hoek}
\author[TAU]{D.~Izraeli}
\author[Mainz]{S.~Kegel}
\author[Mainz]{P.~Klag}
\author[Mainz]{Y.~Kohl}
\author[nrc,TAU]{I.~Korover}
\author[TAU]{J.~Lichtenstadt}
\author[TAU,soreq]{I.~Mardor}
\author[Mainz]{H.~Merkel}
\author[Mainz]{D.G.~Middleton}
\author[UL,JSI,Mainz]{M.~Mihovilovi\v{c} }
\author[Mainz]{J.~M\"uller}
\author[Mainz]{U.~M\"uller}
\author[TAU]{M.~Olivenboim}
\author[TAU]{E.~Piasetzky}
\author[Mainz]{J.~Pochodzalla}
\author[huji]{G.~Ron}
\author[Mainz]{B.S.~Schlimme}
\author[Mainz]{M.~Schoth}
\author[Mainz]{F.~Schulz}
\author[Mainz]{C.~Sfienti}
\author[UL,JSI]{S.~\v{S}irca}
\author[Mainz]{R.~Spreckels}
\author[JSI]{S.~\v{S}tajner }
\author[USK]{S.~Strauch}
\author[Mainz]{M.~Thiel}
\author[Mainz]{A.~Tyukin}
\author[Mainz]{A.~Weber}
\author[TAU]{I.~Yaron}
\author{\\\textbf{(A1 Collaboration)}}

\cortext[cor2]{Corresponding author}

\fntext[ucr]{Present address: UC-Riverside, Riverside, CA 92521, USA.}
\fntext[mit]{Present address: MIT-LNS, Cambridge, MA 02139, USA.}
\address[TAU]{School of Physics and Astronomy, Tel Aviv University, Tel Aviv 69978,
Israel.}
\address[JSI]{Jo\v{z}ef Stefan Institute, 1000 Ljubljana, Slovenia.}
\address[Mainz]{Institut f\"ur Kernphysik, Johannes Gutenberg-Universit\"at, 55099
Mainz, Germany.}
\address[zagreb]{Department of Physics, University of Zagreb, HR-10002 Zagreb, Croatia.}
\address[Rutgers]{Rutgers, The State University of New Jersey, Piscataway, NJ 08855,
USA.}

\address[Pavia]{Dipartimento di Fisica, Universit\`a degli Studi di Pavia and INFN, Sezione di Pavia, via A.~Bassi 6, I-27100 Pavia, Italy.}
\address[Konstanz]{Universit\"at Konstanz, Fachbereich Physik, Universit\"atsstra{\ss}e 10, 78464 Konstanz, Germany.}
\address[nrc]{Department of Physics, NRCN, P.O. Box 9001, Beer-Sheva 84190, Israel.}
\address[soreq]{Soreq NRC, Yavne 81800, Israel.}
\address[huji]{Racah Institute of Physics, Hebrew University of Jerusalem, Jerusalem
91904, Israel.}
\address[UL]{Faculty of Mathematics and Physics, University of Ljubljana, 1000
Ljubljana, Slovenia.}
\address[USK]{University of South Carolina, Columbia, South Carolina 29208, USA.}

\begin{abstract}
We report measurements of the induced polarization $\vec P$ of protons knocked out from $^2$H and $^{12}$C via the $A(e,e'\vec p\,)$ reaction.  We have studied  the dependence of $\vec P$ on two kinematic variables: the missing momentum $p_{\rm miss}$ and the ``off-coplanarity'' angle $\phi_{pq}$ between the scattering and reaction planes.  For the full 360$\degree$ range in $\phi_{pq}$, both the normal ($P_y$) and, for the first time, the transverse ($P_x$) components of the induced polarization were measured with respect to the coordinate system associated with the scattering plane.  $P_x$ 
vanishes in coplanar kinematics, however in non-coplanar kinematics, it is on the same scale as $P_y$.

We find that the dependence on $\phi_{pq}$ is sine-like for $P_x$ and cosine-like for $P_y$.  
For carbon, the magnitude of the induced polarization is especially large when protons are knocked out from the $p_{3/2}$ shell at very small $p_{\rm miss}$.  For the deuteron, the induced polarization is near zero at small $|p_{\rm miss}|$, and its magnitude increases with $|p_{\rm miss}|$.  For both nuclei such
behavior is reproduced qualitatively by theoretical results, driven largely by the spin-orbit part of the final-state interactions. However, for both nuclei, sizeable discrepancies exist between experiment and theory.

\end{abstract}

\end{frontmatter}{}

\section{Introduction}

Within the shell model a spin-orbit term is required in the mean-field potential
of atomic nuclei in order to explain the energy splitting of the single-particle
levels for reproducing the magic numbers
 \cite{PhysRev.78.16,SORLIN2008602,PhysRev.75.1766.2}.  The spin-orbit interaction also plays an important role in optical potentials which describe scattering processes.   These have a strong influence in the final-state interactions (FSI) affecting quasi-free $A(e,e'\vec p\,)$ scattering \cite{BOFFI1980437,Giusti:1989ww,Boffi:1996ikg,Meucci:2001qc,PhysRevC.70.044608}, as well as various other types of scattering processes 
  \cite{PhysRevLett.12.108,PhysRev.147.829,HORIKAWA1980386,WATANABE1958484}.

In elastic $ep$ scattering, within
the one-photon-exchange approximation, the induced
polarization of the proton vanishes. Consequently, it is the FSI
which in the $A(e,e'\vec p\,)$ reaction gives rise to a non-vanishing
induced polarization of the knocked-out proton.  
In view of the fact that it is largely insensitive to details of the nucleon electromagnetic form factors, the induced polarization serves as  an effective probe of FSI effects in quasi-elastic $A(e,e'\vec p\,)$. 
Here we present measurements of the induced polarization of quasi-elastic protons from $^{2}$H and $^{12}$C over a wide range in the missing momentum, $p_{\rm miss}$.

Previous measurements of the normal component, $P_y$, of the induced polarization have been performed at MIT-Bates on $^2$H \cite{Milbrath:1997de} at low $p_{\rm miss}$  and $^{12}$C  with large $p_{\rm miss}$ in coplanar kinematics  \cite{Woo}.  Measurements of $P_y$ were also performed on $^4$He at Jefferson Lab (JLab) \cite{PhysRevLett.91.052301,Malace_2011}  over a wide $p_{\rm miss}$ range.

In Ref.~\cite{Woo}, it was found that the induced polarizations of protons knocked out from the $s$ shell of $^{12}$C show a different behavior from those knocked out of the $p$ shell.  This difference was attributed to the spin-orbit ($L\cdot S$) interaction.  
The measured values of $P_y$ for $^2$H at low $p_{\rm miss}$ in \cite{Milbrath:1997de} were much smaller than those measured for other nuclei in \cite{Woo,PhysRevLett.91.052301,Malace_2011}.  All measurements prior to those reported here were restricted to almost-coplanar geometry.

The induced polarization measurements presented here were performed at the Mainz Microtron (MAMI), during four run periods from 2012-2017.
  Our measurements for both nuclei cover a large range in missing momentum, $p_{\rm miss}$, and the full 360$\degree$ range in the off-coplanarity angle, $\phi_{pq}$ (See Fig.~\ref{fig:kinematics_diagram}).  Calculations predict a dependence of the induced polarization on $\phi_{pq}$ largely due to the $L\cdot S$ interaction which hitherto has been unexplored.   Our $^2$H data greatly extend the range in $p_{\rm miss}$ compared to the previous measurements in \cite{Milbrath:1997de}, while our $^{12}$C data significantly improve the statistical precision and the range in $\phi_{pq}$ compared to the existing data \cite{Woo}.  Furthermore, we measure not only the normal component, $P_y$, but also, for the first time, the transverse component, $P_x$, which vanishes in coplanar kinematics ($\phi_{pq} = 0\degree$ or $180\degree$).  
The \textit{transferred} polarizations measured in our experiments were reported in \cite{deep2012PLB,deepCompPLB,posPmissPLB} for $^{2}$H and \cite{ceepLet,ceepComp,ceepTim} for $^{12}$C.

Section \ref{sec:setup} describes the experimental setup, the measured reaction, and the kinematic settings.   
The data analysis and extraction of the induced polarization are described in Sec.~\ref{sec:measurement}.  The details of theoretical calculations, to which we compare our data, are given in Sec.~\ref{sec:calc}.  We then present the data for both nuclei and their dependence on $p_{\rm miss}$ and $\phi_{pq}$ in Secs.~\ref{sec:pmiss} and \ref{sec:phipq}, and conclude in Sec.~\ref{sec:conclusions}. 

\section{Experimental setup and kinematics}
\label{sec:setup}
The experiments were performed at MAMI using the A1 beamline and spectrometers \cite{a1aparatus}.  For these measurements, a 600-690 MeV polarized continuous-wave electron beam was used.  The beam current was $\approx$10 $\mu$A.  Due to the frequent flipping of the beam helicity (about 1 Hz), the average beam polarization in our event sample is zero, as verified by internal checks on the data.

 The targets used for the $^2$H and $^{12}$C measurements were a 50 mm long oblong cell filled with liquid deuterium \cite{deep2012PLB,deepCompPLB,posPmissPLB} and a set of three 0.8 mm-thick graphite foils \cite{ceepLet,ceepComp,ceepTim}, respectively.  We also performed calibration runs using a liquid hydrogen target.

 Two high-resolution, small-solid-angle spectrometers with momentum acceptances of 20-25\% were used to detect the scattered electrons and knocked-out protons in coincidence.  Each of these spectrometers consists of an momentum-analysing magnet system followed by a set of vertical drift chambers (VDCs) for tracking, and a scintillator system for triggering and defining the time coincidence between the two spectrometers.

 The proton spectrometer was equipped with a focal-plane polarimeter (FPP)  with a 3-7 cm thick carbon analyzer and a set of horizontal drift chambers (HDCs) \cite{a1aparatus,Pospischil:2000pu}. The spin-dependent scattering of the polarized proton by the carbon analyzer allows the determination of the proton polarization at the focal plane.  The polarization at the interaction point is then determined by correcting for the spin precession  in the spectrometer's magnetic field \cite{Pospischil:2000pu}. 
 More details of the experiment can be found in \cite{deep2012PLB,deepCompPLB,posPmissPLB,ceepLet,ceepComp,ceepTim}.

\begin{figure}[ht]
\includegraphics[width=\columnwidth]{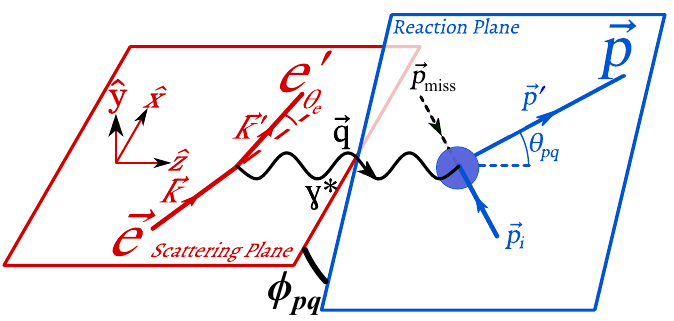}
\caption{
Kinematics of the reaction with the definitions of the kinematic variables.
}
\label{fig:kinematics_diagram}
\end{figure}

\begin{table*}[h!]
\caption{
The kinematic settings in the $^{2}{\rm H}(\vec{e},e'\vec p\,)$ and $^{12}{\rm C}(\vec{e},e'\vec p\,)$ measurements. The angles and momenta represent the central values for the two spectrometers: $p_p$ and $\theta_p$ ($p_e$ and $\theta_e$) are the knocked out proton (scattered electron) momentum and scattering angles, respectively.  The number of events passing the event selection cuts are also given.  
}
\begin{center}
\begin{tabular}{lllllllll}
\hline
\multicolumn{2}{c}{}& \multicolumn{7}{l}{Kinematic setting}  \\ \cline{3-9}
 & & A & B & C & D &E & F & G\\

\hline
$E_{\rm beam}$ & [MeV] & 600  & 600 & 630 & 630 & 690 & 690 & 600\\
$Q^2$ & [$({\rm GeV}\!/\!c)^2$] & 0.40 & 0.40 & 0.18 & 0.18  & 0.65 & 0.40 & 0.18\\
$p_{\rm miss}$ & [MeV$\!/\!c$]  & $-$80 to 75 & 75 to 175 & $-$80 to $-15$ & $-$220 to $-$130 & 60 to 220 & $-$70 to 70 & $-$250 to $-$100\\
$p_e$ & [MeV$\!/\!c$] & 384 & 463 & 509 & 398 & 464 & 474 & 368 \\
$\theta_e$ & [deg] & 82.4 & 73.8 & 43.4 & 49.4 & 90.9  & 67.1 & 52.9\\
$p_p$ & [MeV$\!/\!c$] & 668 & 495 & 484 & 665 & 656  & 668 & 665\\
$\theta_{p}$ & [deg] & $-$34.7 & $-$43.3 & $-$53.3 & $-$39.1 & $-$33.6 & $-$40.8 & -37.8\\
\hline
\\
Nucleus & shell & \multicolumn{3}{l}{\small \# of events passing cuts ($\times 10^3$)}\\
\hline
 $^2{\rm H}$ & & 68  & 19 & 438 & 201 & 10 & 232 & --- \\
 $^{12}{\rm C}$ & $s_{1/2}$ & 268  &---&---&---&---&---&274 \\
 $^{12}{\rm C}$ & $p_{3/2}$ & 160 &---&---&---&---&---& 436 \\
\hline
\end{tabular}

\end{center}
\label{tab:kinematics_D}
\end{table*}
 The kinematics of the measured reactions are shown in Fig.~\ref{fig:kinematics_diagram}.  The electron's initial and final  momenta are $\vec k$ and $\vec k\,'$ respectively, which define the scattering plane of the reaction.  The reaction plane is defined by the momentum transfer $\vec q = \vec k-\vec k\,'$ and the recoiling proton's momentum $\vec p\,'$.  We refer to the angle between the scattering plane and the reaction plane as the ``off-coplanarity'' angle of the reaction, denoted by $\phi_{pq}$.  
 
 Following the convention of \cite{PhysRevLett.91.052301}, we express the components of the induced polarization $\vec P$ in the scattering-plane coordinate system, such that $\hat y$ is normal to the scattering plane (along the direction of $\vec k \times \vec k\,'$), $\hat z$ is along the direction of the momentum transfer $\vec q$, and $\hat x = \hat y\times\hat z$, forming a right-handed coordinate system.

The missing momentum  $\vec p_{\rm miss}\equiv \vec q - \vec p\,'$ is the recoil
momentum of the residual nuclear system. Neglecting FSI, 
$-\vec p_\mathrm{miss}$ is equal to the initial momentum of the emitted proton, $\vec p_i$.
We conventionally define positive and negative signs for $p_{\rm miss}$ by the sign of $\vec p_{\rm miss}\cdot\vec q$.

Our $^2$H measurements were performed at six kinematic settings, labeled A through F, with varying ranges of $p_{\rm miss}$ and invariant four-momentum transfers $Q^2=-q^2$.  Settings A and F were both centered at $p_{\rm miss} = 0$, and have $Q^2 = 0.40~({\rm GeV}\!/\!c)^2$.  Settings B and E covered large positive $p_{\rm miss}$, at $Q^2 = 0.40$ and 0.65 $({\rm GeV}\!/\!c)^2$, respectively.
Settings C and D covered small and large negative $p_{\rm miss}$, respectively, and were both at $Q^2 = 0.18~({\rm GeV}\!/\!c)^2$.  Details are given in Table \ref{tab:kinematics_D}.

Our $^{12}$C measurements were taken at two kinematic settings.   The first is the same Setting A of the deuteron measurements (centered near $p_{\rm miss} = 0$, at $Q^2 = 0.40~({\rm GeV}\!/\!c)^2$).  The second is Setting G, which covered a region of large negative $p_{\rm miss}$ at $Q^2 = 0.18~({\rm GeV}\!/\!c)^2$; this is similar to Setting D of the deuteron measurements, except with a different beam energy and the other kinematic variables modified accordingly\footnote{In our earlier publications \cite{ceepLet,ceepComp,ceepTim} on $^{12}$C, this setting is referred to as Setting B.~  We refer to this setting as Setting G in this work in order to distinguish it from the Setting B of the deuteron measurements.}. 

In each of the kinematic settings presented in this work, the spectrometers' reference trajectories form a parallel reaction ($\vec p\,' \parallel \vec q\,$).  However, due to the spectrometer acceptance, our data sample included reactions with $\theta_{pq}$ (the angle between $\vec p$ and $\vec q\,$) up to $\approx 8\degree$, with the full 360$\degree$ range in the off-coplanarity angle $\phi_{pq}$.

\section{Data Analysis}
\label{sec:measurement}
\subsection{Event selection}
\label{sec:selection}
Software cuts were applied to the data, in order to ensure good tracking, time coincidence, and event quality.  These cuts applied here are identical to those of the earlier publications \cite{deep2012PLB,deepCompPLB,posPmissPLB,ceepLet,ceepComp,ceepTim} on the transferred polarization, unless otherwise noted below.  

We applied additional tracking cuts to the proton's trajectory, requiring it to be within the part of the spectrometer where the precession of the proton's spin is well known and the false asymmetry could be determined using dedicated elastic $ep$ measurements.  
In order to reduce false asymmetries, we also removed events where the proton would either be outside of the geometric acceptance of the detector, or produce a hit on a malfunctioning channel of the HDCs, if it had scattered in the azimuthally opposite direction.
Additionally, the polar angle $\Theta_{\rm FPP}$  of the secondary scattering was required to be greater than $8\degree$ in order to avoid spin-independent Coulomb-scattering events, and less than 
$23\degree$ 
 in order to improve the stability of the false-asymmetry determination.

Following \cite{deep2012PLB,deepCompPLB,posPmissPLB}, we required the missing mass of the $^2$H$(e,e'\vec p\,)$ reaction to be consistent with the mass of a neutron. 
For the $^{12}$C sample, we distinguish between protons knocked out from the $s$ and $p$ shells, following \cite{ceepLet,ceepComp,ceepTim}, by using cuts on the missing energy, $E_{\rm miss}$ in the reaction, defined as  \cite{Dutta}:

\begin{equation}
E_{\rm miss} \equiv \omega - T_p - T_{^{11}\rm B},
\end{equation}
 where $\omega=k^0-k'^{\,0}$ is the energy transfer, $T_p$ is the measured kinetic energy of the outgoing proton, and $T_{^{11}\rm B}$ is the calculated kinetic energy of the recoiling residual system, assuming it is $^{11}$B in the ground state.  
    For the $s$-shell sample, we used the cut $30< E_{\rm miss} < 60$ MeV,  while for the $p$-shell sample, we used  $15< E_{\rm miss} < 25$ MeV \cite{ceepLet,ceepComp,ceepTim}.
 
 The $p$-shell cut accepts events in which the residual $A-1$ system is left in one of several discrete states, including the ground-state of $^{11}$B as well as a few excited states.  
 The $s$-shell selection cut is much wider, comprising a broad range within the continuum of unbound residual $A-1$ states.  

\subsection{Polarization fitting}
\label{subsec:fitting}
Before extracting the values of $P_x$ and $P_y$ for $^2$H and $^{12}$C, we first determined the false asymmetry using elastic $ep$ events (for which the induced polarization is expected to be zero).  This was accomplished by maximizing the log likelihood
\begin{equation}
\log \mathcal{L} = \sum\limits_{\rm events} \log\left[1+\vec A^{\,\rm T} \cdot\left(\begin{array}{c}-\sin{\Phi_{\rm FPP}} \\\cos{\Phi_{\rm FPP}} \\0\end{array}\right)\right],
\end{equation}
for the $ep$ event sample, where $\Phi_{\rm FPP}$ is the azimuthal angle of the secondary scattering and $\vec A$ is  the false asymmetry in the focal plane coordinate system, parameterized as
\begin{equation}
\vec A = \left(\begin{array}{c}a^x_{0} + a^x_{1}\phi_{\rm vth}\\a^y_{0} + a^y_{1}\theta_{\rm vth}  \\0\end{array}\right),
\end{equation}
where $\theta_{\rm vth}$ and $\phi_{\rm vth}$ are the incident angles of the proton trajectory extrapolated from the VDCs to the HDCs.  $a^x_0$, $a^x_1$, $a^y_0$, and $a^y_1$ are the fitted coefficients.
We then extracted the induced polarization for $^2$H and $^{12}$C by maximizing the log likelihood 
\begin{equation}
\log \mathcal{L} = \sum\limits_{\rm events} \log\left[1+(a\mathbf{S}\cdot \vec P+\vec A)^{\rm T}\cdot\left(\begin{array}{c}-\sin{\Phi_{\rm FPP}} \\\cos{\Phi_{\rm FPP}} \\0\end{array}\right)\right],
\end{equation}
where  $\mathbf{S}$ is the calculated spin-transfer matrix for the proton trajectory of the event, and $a$ is the analyzing power of the event (as determined by \cite{AprileGiboni:1984pb,Mcnaughton:1986ks}).  $\vec P$ is the induced polarization.  We constrain $P_z$ to be zero in order to improve the stability of our fit.  This constraint has a negligible effect on the fitted $P_x$ and $P_y$ except in bins with very poor statistics.  

The corrections for false asymmetry are larger for  $P_x$ than for $P_y$; the r.m.s. values of these corrections are $\approx$0.20 and $\approx$0.04 respectively.  This is because $|A_y|$ is generally larger than $|A_x|$, and the off-diagonal terms of the spin-transfer matrix, $S_{xy}$ and $S_{yx}$, dominate over the much smaller diagonal terms $S_{xx}$ and $S_{yy}$.  Details of the false-asymmetry determination, and the checks we used to validate its long-term stability, may be found in the supplementary material.  

\subsection{Systematic errors}
\label{sec:systematics}
The systematic
errors in these measurements are due to a few sources, which are presented in Table \ref{tab:systematics}.   They are dominated by the uncertainty on the false asymmetry of the FPP.~ This is due to the limited statistics of the elastic $ep$ sample used to determine this false asymmetry, and it contributes 0.012 to the systematic error on our corrected results for $^2$H and $^{12}$C.  

The analyzing
power of the carbon secondary scatterer is known to about 1\%  in this kinematic region \cite{Pospischil:2000pu,AprileGiboni:1984pb,Mcnaughton:1986ks}. It leads to a relative error of the same size on each component of $\vec P$. The uncertainty on the precession of the proton's spin introduces an additional 0.4\% relative error.  The systematic error due to the uncertainty of the alignment between the HDC and the VDC detector systems was investigated to be less than 0.001, absolute, for both components.  This was determined by repeating the analysis with each of the alignment parameters modified by plus or minus its uncertainty.  In a similar manner, we estimate the uncertainty on both components due to the uncertainty on the kinematic settings (i.e. the beam energy and the two spectrometers' angles and momenta)  to be about 0.001. 

The software cuts, described in Sec.~\ref{sec:selection}, introduce an additional absolute uncertainty of $\approx$0.006 to the overall systematic error.  This was determined by performing slightly tighter cuts on each of the software cuts and then extracting the polarization as described above.  The systematic uncertainty due to the cuts is then the quadrature sum of the deviations between the measured polarizations with each of the tightened cuts as compared to the standard set of cuts.  

\begin{table}[h]
\caption{Sources of systematic errors on $P_x$ and $P_y$.  We distinguish between sources of systematic errors that do not scale with $P_y$ (absolute errors), and those that do (relative errors).  The total systematic errors are then $\Delta P_x = \sqrt{\Delta P^2_{x,\rm abs}+(\Delta P_{x,\rm rel} P_x)^2}$ and similarly for $\Delta P_y$.}

\begin{tabular}{lrrr}
\hline
 & $\Delta P_{x,\rm abs}$ & $\Delta P_{y,\rm abs}$ & $\Delta P_{xy,\rm rel}$\\
\hline
False asymmetry & 0.010 & 0.012 & ---   \\
Software cuts & 0.005 & 0.006 & --- \\
Detector alignment & $<0.001$ & $<0.001$ & --- \\
Kinematic setting & $0.001$ & $0.001$ & --- \\
Precession & --- & --- & 0.4\% \\
Analyzing power & --- & --- & 1.0\% \\
\hline
Total & 0.011& 0.013 & 1.1\%\\
\hline
\end{tabular}

\label{tab:systematics}
\end{table}

\begin{figure}[h!]

\begin{center}
{\huge \textrm{$^2{\rm H}(e,e'\vec p\,)$}}
\includegraphics[width=\columnwidth]{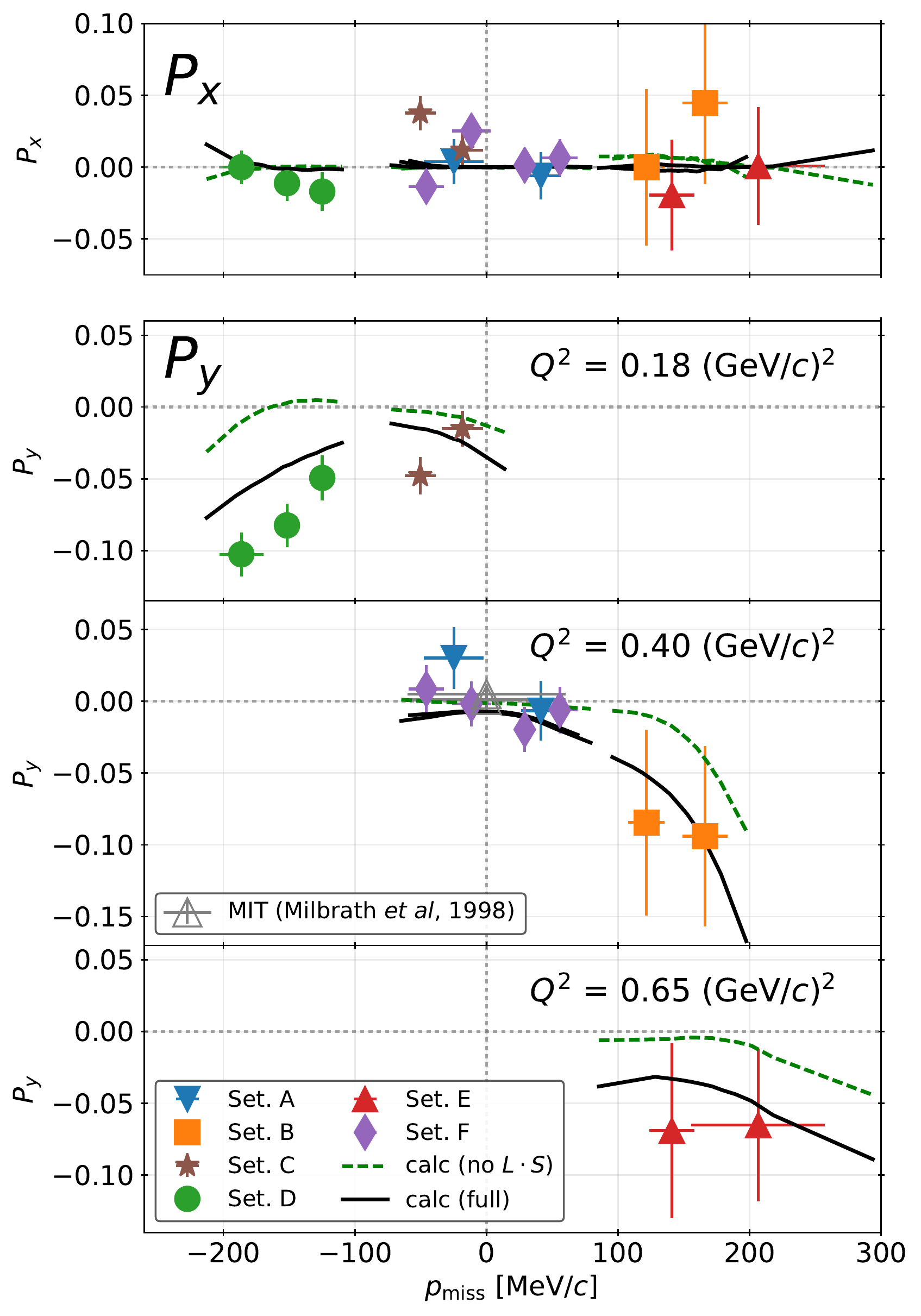}
\end{center}
\caption{
The measured components of the induced polarization for $^2$H,  $P_x$ (top panel) and $P_y$ (three lower panels for each $Q^2$), as functions of the missing momentum. Different symbols (color online) represent different kinematic settings as shown in the inset of the lowest panel.  Grey open triangles indicate the measurements from MIT-Bates \cite{Milbrath:1997de}, taken at $Q^2$ = 0.38 and 0.50 $({\rm GeV}\!/\!c)^2$, both with $p_{\rm miss}$ centered at zero. The uncertainties shown are statistical only. Systematic errors are discussed in Sec.~\ref{sec:systematics}. Theoretical results with (without) the $L\cdot S$ part of the interaction are shown as solid black (dashed, green online) curves. 
}
\label{fig:deuteron_pmiss}
\end{figure}

\begin{figure*}[h]

\begin{center}
{\huge $^{12}{\rm C}(e,e'\vec p\,)$}

\includegraphics[width=\columnwidth]{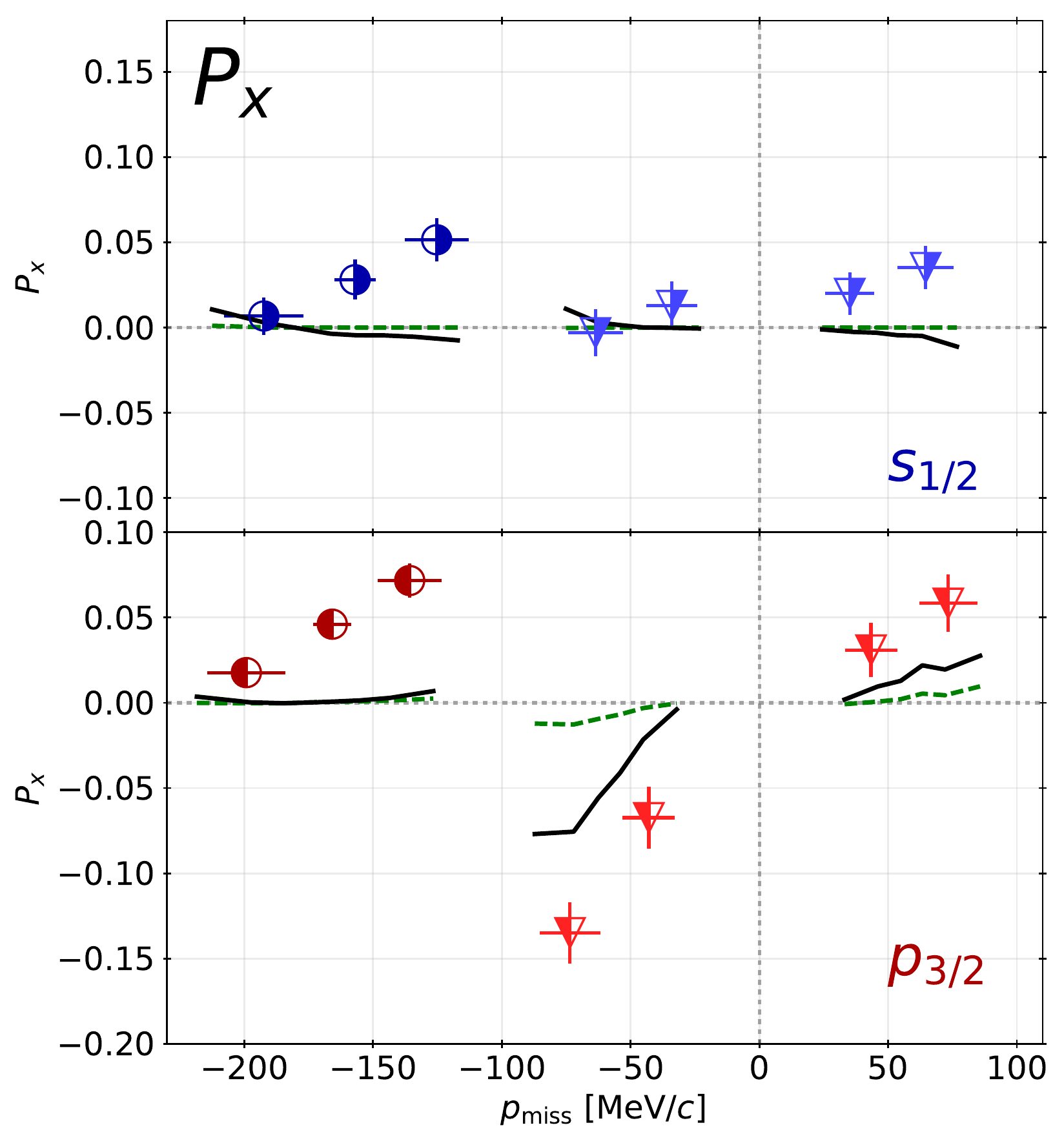}\hspace{7mm}
\includegraphics[width=\columnwidth]{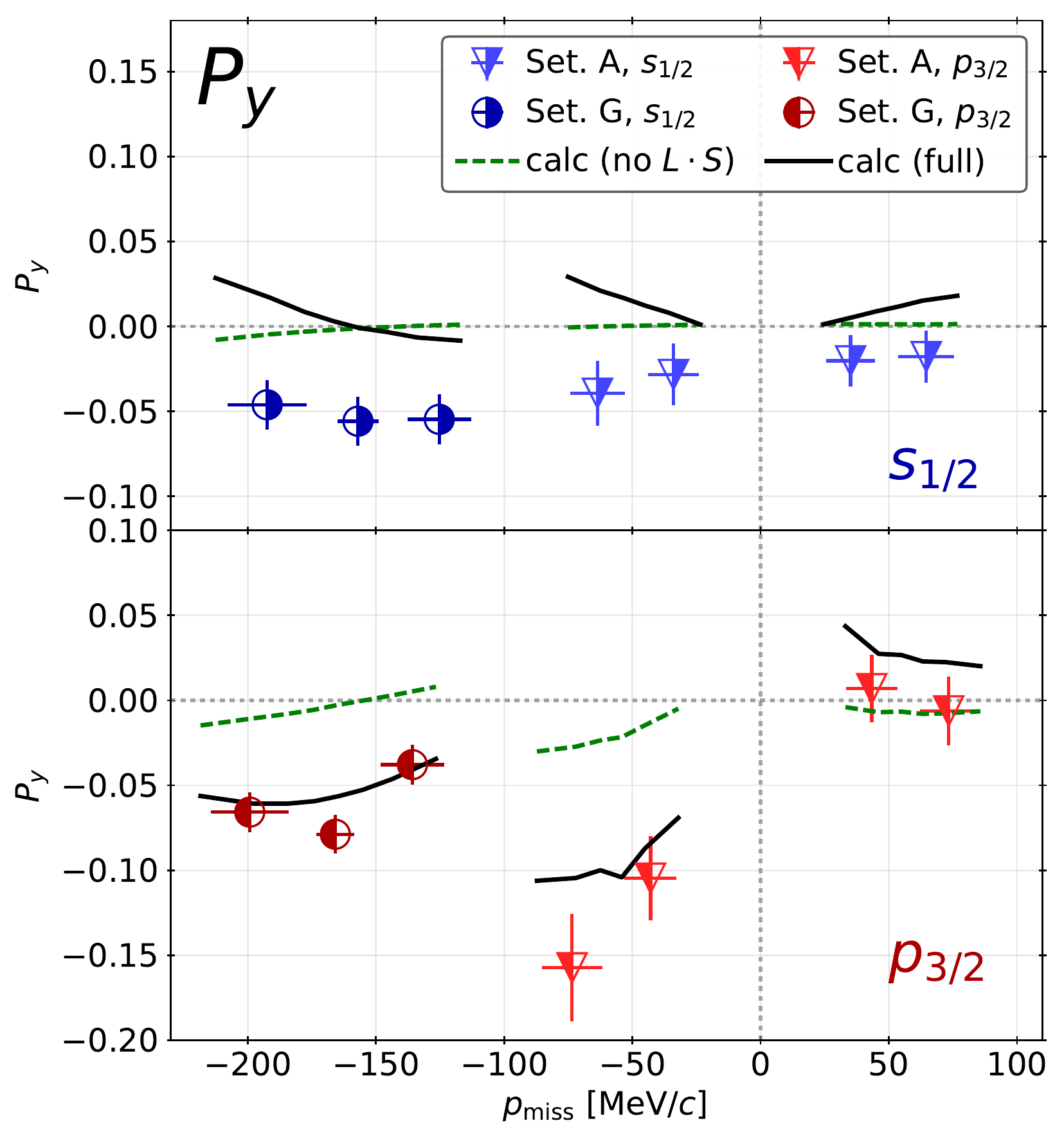}
\end{center}

\caption{ 
The measured induced-polarization components, $P_x$ (left panel) and $P_y$ (right panel) for $^{12}$C as functions of the missing momentum compared to theory. Triangles (circles) refer to kinematic
Setting A (G).~ Symbols that are open on the left (right) side refer to
$s$-shell ($p$-shell) removals, and are colored blue (red) online.
The calculations with (without) the $L\cdot S$ potential are shown in solid
black (dashed, green online) curves.
}
\label{fig:carbon_pmiss}
\end{figure*}

\section{Calculations}
\label{sec:calc}
For comparison, theoretical calculations of the induced
polarization for $^2$H and $^{12}$C have been performed.
For $^2$H we have used a non relativistic calculation \cite{Arenhovel} including
a realistic $NN$-potential, meson-exchange (MEC) and isobar (IC)
currents, and relativistic contributions (RC) of leading order.
For the bound and scattering states the realistic Argonne $V_{18}$ 
potential \cite{PhysRevC.51.38} has been taken.  As nucleon electromagnetic form factors
      we used the parameterizations from  \cite{Bernauer}.

For $^{12}$C, calculations were performed using a program \cite{Meucci:2001qc} based on the relativistic distorted-wave approximation (RDWIA) 
where the FSI between the outgoing proton and the residual nucleus are described
by a phenomenological relativistic optical potential.  
The original program \cite{Meucci:2001qc} was modified \cite{ceepTim} in order to account for non-coplanar kinematics, by including all relevant structure functions \cite{Boffi:1996ikg}.  In the RDWIA calculations, only the
one-body electromagnetic nuclear current is included. We chose the
current operator corresponding to the cc2 definition \cite{DEFOREST1983232},
and we used the same
parametrization of the nucleon form factors \cite{Bernauer} as in the $^2$H
calculations. The relativistic  proton bound-state wave functions were
obtained from the NL-SH parametrization  \cite{SHARMA1993377} and the scattering states
from the so-called  ``democratic'' parameterization of the optical
potential \cite{PhysRevC.80.034605}

We found that for both nuclei, the calculated induced polarization has very little sensitivity to the details of the nucleon form factors.  
A change of 10\% to the form-factor ratio $G_E/G_M$ in the calculations affects the induced polarization by less than 0.005. 

In order to examine the influence of the $L\cdot S$ interaction
on the induced polarization, we repeated these calculations while switching off
this part of the potential.
As we will show in Secs.~\ref{sec:pmiss} and \ref{sec:phipq}, the $L\cdot S$ interaction is the dominant source of the induced polarization.  

\begin{figure*}[h!]

\begin{center}
{\huge $^2{\rm H}(e,e'\vec p\,)$}

\includegraphics[width=\columnwidth]{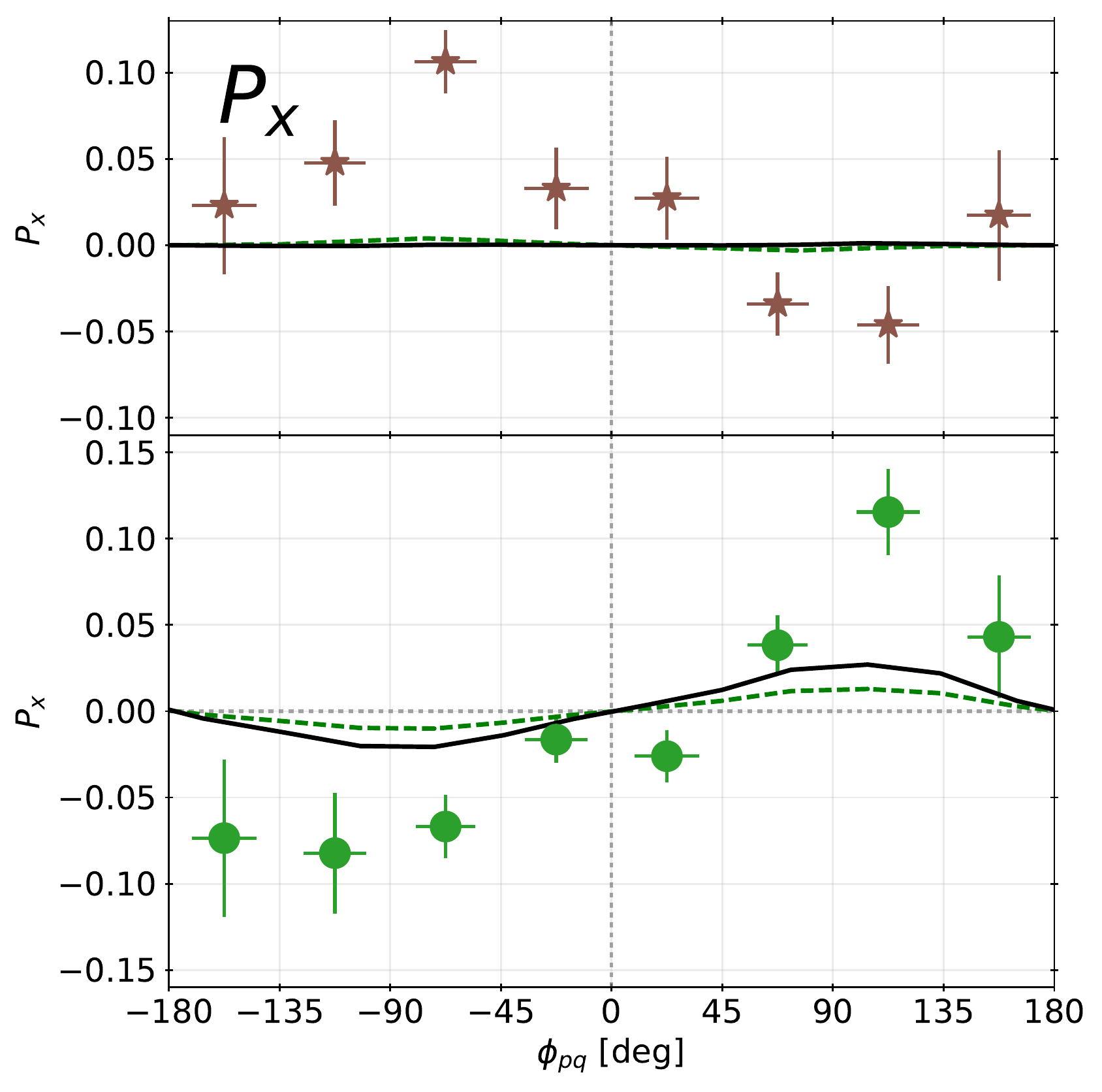}\hspace{7mm}
\includegraphics[width=\columnwidth]{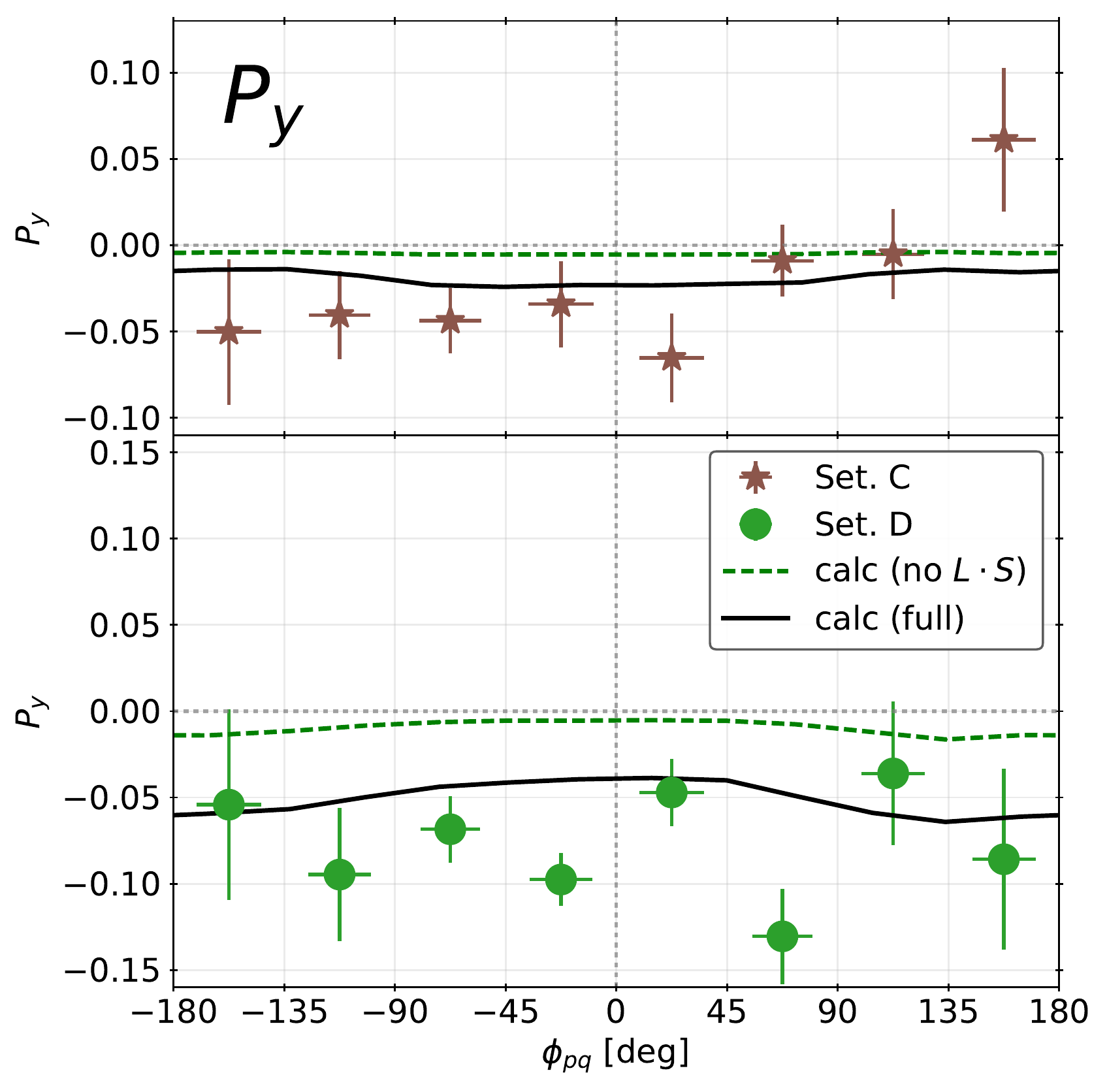}
\end{center}

\caption{
Comparison of the $\phi_{pq}$ dependence of the measured induced
polarization components $P_x$ (left panels) and $P_y$ (right panels) for $^2$H
to the theory for selected kinematic settings. Symbols are explained in
the inset of the right bottom panel. 
The calculations with (without) the $L\cdot S$ potential are shown as the solid
black (dashed, green online) curves.
}
\label{fig:deuteron_alpha_select}
\end{figure*}
\section{Missing-momentum dependence}
\label{sec:pmiss}

We partitioned the $^2$H data for each kinematic setting into bins by $p_{\rm miss}$ and extracted $P_x$ and $P_y$ for each bin separately, and present the results in Fig.~\ref{fig:deuteron_pmiss}.  The results for $P_x$ (top panel) are consistent with zero within error.  The extracted values of $P_y$, on the other hand, are near zero at small $p_{\rm miss}$ and deviate from zero at large $|p_{\rm miss}|$. 
This is consistent with earlier measurements of $P_y$ from MIT \cite{Milbrath:1997de} (grey pentagons in Fig.~\ref{fig:deuteron_pmiss}), which show that for $|p_{\rm miss}| < 60$ MeV/$c$, the values of $P_y$ are consistent with zero.  The reason for this feature lies in the fact that for $p_{\rm miss} = 0$,
     which is the ideal quasi-free case, the influence of FSI is
     very small, almost zero.

 In order to compare with theory, we calculated $P_x$ and $P_y$ for a sample of the $^2$H events which reflect the distribution of events in the phase space at each kinematic setting and took the average value by bins in $p_{\rm miss}$, as shown in Fig.~\ref{fig:deuteron_pmiss} by solid black curves.  The calculations for $P_x$ vanish, while those for $P_y$ are near zero at $p_{\rm miss}\approx0$, but they become increasingly negative at increasing $|p_{\rm miss}|$.  The $P_y$ calculations match the data very well except at large negative $p_{\rm miss}$ (Setting D).

The theoretical results obtained by switching off the $L\cdot S$
interaction are presented as dashed (green online) curves  in Fig.~\ref{fig:deuteron_pmiss}.
Similar to the complete calculation, the calculated $P_x$ without the $L\cdot S$ interaction is essentially
zero. For $P_y$, the corresponding difference between the full and no-$L\cdot S$
calculation is large, indicating the important influence of the $L\cdot S$
interaction on $P_y$. The theory appears to describe the data
quite well except for $P_y$ in Settings C and D.

\begin{figure*}[t]

\begin{center}
{\huge $^{12}{\rm C}(e,e'\vec p\,)$}

\includegraphics[width=\columnwidth]{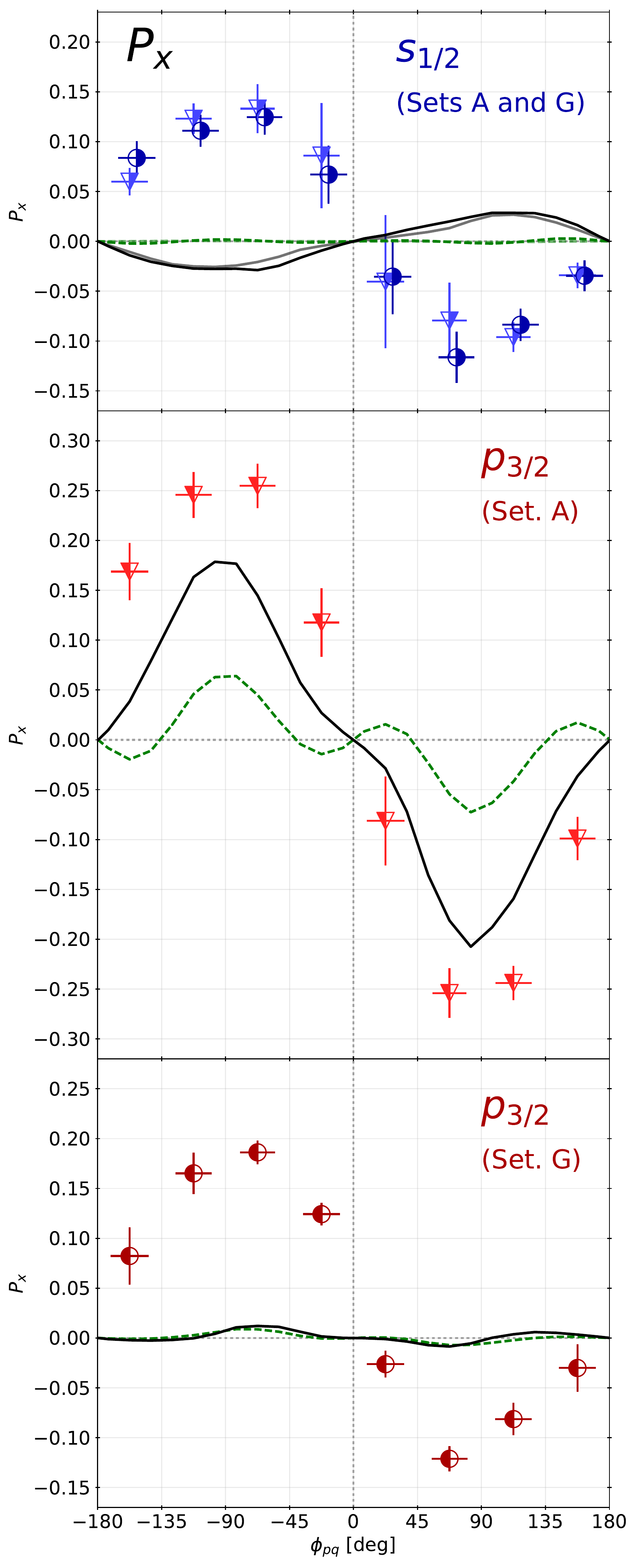}\hspace{7mm}
\includegraphics[width=\columnwidth]{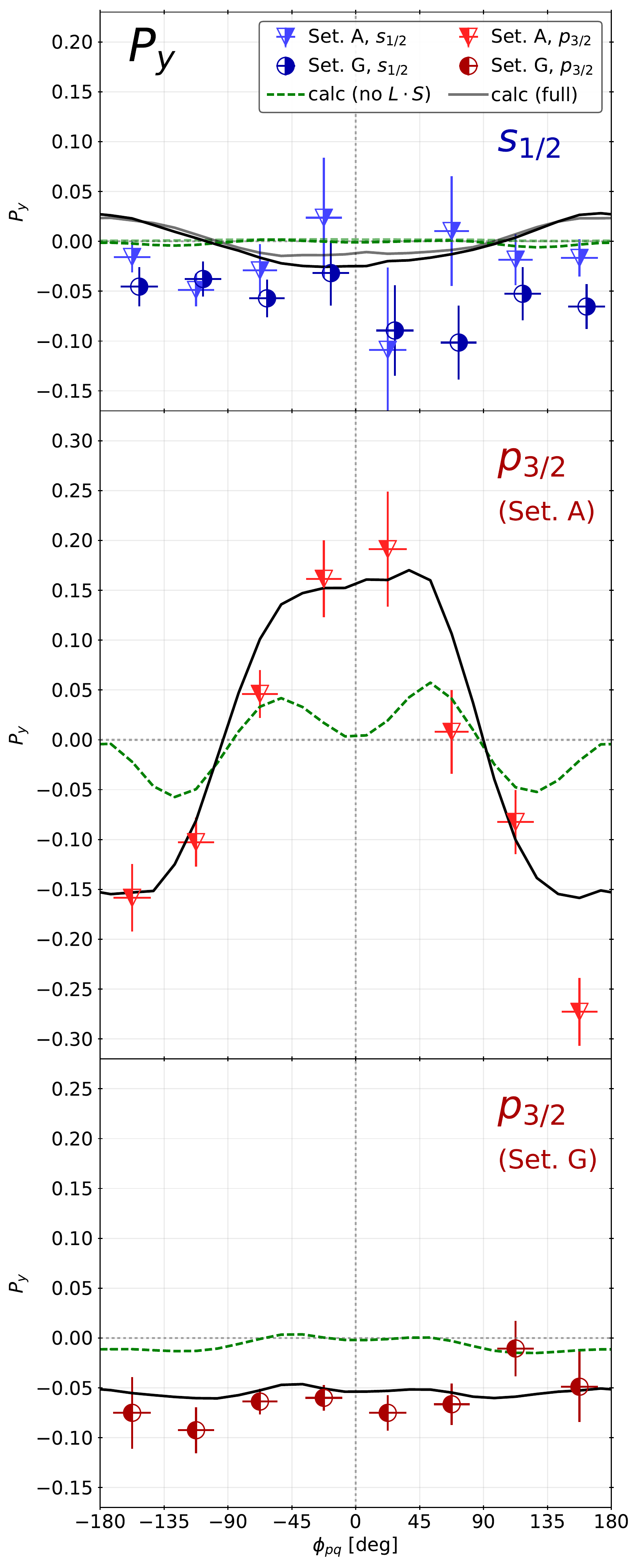}
\end{center}
\caption{
For $^{12}$C, the $\phi_{pq}$ dependence of the measured polarization components
$P_x$ (left panels) and $P_y$ (right panels) compared to theory.
These are shown for $s$-shell knockout for both kinematic settings (top panels),
and for $p$-shell knockout at small $p_{\rm miss}$ (Setting A, middle panels) and large
negative $p_{\rm miss}$ (Setting G, bottom panels). Calculations with (without)
the $L\cdot S$ part of the optical potential are shown as solid black (dashed, green online) curves.
In the top panels, the curves for Setting A are shown in grey (light green online for no-$L\cdot S$) in
order to contrast with those of Setting G.
}
\label{fig:carbon_alpha}
\end{figure*}

For $^{12}$C we performed the same polarization-fitting
procedure and present the results in Fig.~\ref{fig:carbon_pmiss}.
  Unlike in our results for $^2$H, the measured $P_x$ for $^{12}$C is non-zero when binned in $p_{\rm miss}$. 
While a non-zero $P_x$ at coplanar kinematics is theoretically forbidden (see \cite{Giusti:1989ww}), our data are not restricted to coplanarity, and therefore we do not require $P_x$ to be zero.  The non-zero values of $P_x$ in the left panels of Fig.~\ref{fig:carbon_pmiss} reflect the distributions of other kinematic variables (such as $\phi_{pq}$) within each $p_{\rm miss}$ bin. We will discuss this in further detail in Sec.~\ref{sec:phipq}, where we will show that $P_x$ is highly dependent on $\phi_{pq}$.

The component $P_y$ for the $s$-shell data  (upper part of the
right panel of Fig.~\ref{fig:carbon_pmiss}) appears to be nearly constant
at $P_y\approx -0.04$ as function of $p_{\rm miss}$,
while the $p$-shell data show a significant variation with $p_{\rm miss}$.

A comparison of our $^{12}$C data for $P_y$ with the earlier MIT measurements \cite{Woo} is shown in the supplementary material.   We found that our data are consistent with the latter when restricting the range of $\phi_{pq}$ in our data sample to match that of \cite{Woo} (i.e., near $180\degree$) and comparing them at the same $|p_{\rm miss}|$.

The calculations for the $^{12}$C events using 
samples of $s$-shell and p-shell data from both kinematic sets,
with (without) the $L\cdot S$ part of the optical potential, as 
functions of the binned $p_{\rm miss}$, are shown in Fig.~\ref{fig:carbon_pmiss} as solid black (dashed, green online) curves. The results including the $L\cdot S$ term show
much larger deviations from zero than the results without it,
indicating that most of the deviation from zero originates
from the $L\cdot S$ term.

The $P_x$ curves are non-zero, and reflect the asymmetries of the kinematics of the accepted events.  
For $P_y$, the full-potential calculation (solid black) curves are above the measured data points by up to 0.05, and in some cases, the calculations have the opposite sign of the measured $P_y$.

\section{Dependence on off-coplanarity}
\label{sec:phipq}
In order to examine the $\phi_{pq}$ dependence of $P_x$ and $P_y$, we performed the fits to the data in bins of $\phi_{pq}$.  

The results for $^2$H at the negative-$p_{\rm miss}$ kinematic settings (C and D) are shown in Fig.~\ref{fig:deuteron_alpha_select}.  These are the only two settings for which we observe a significant $\phi_{pq}$ dependence in the data (for completeness, the results for the other settings are shown in the supplementary material).  In both of these two settings, the $P_x$ have a sine-shaped dependence on $\phi_{pq}$, albeit with opposite signs for the two settings.  In all six of the $^2$H settings, $P_y$ shows no statistically significant variation with respect to $\phi_{pq}$.

\FloatBarrier
The theory predicts for $P_x$ for Setting C an almost vanishing $\phi_{pq}$ dependence in contrast to the data, while for Setting D a small sine-shaped behavior is obtained, however, with a much smaller amplitude than the much more pronounced data. For $P_y$ the agreement with the data appears to be better.   For both settings, there is only a small variation in $P_y$ with respect to  $\phi_{pq}$, and its amplitude is larger for Setting D than for Setting C.  
Furthermore, one notes for both components a sizeable contribution for the case where the $L\cdot S$-part is switched-off.
\FloatBarrier

 The results for $^{12}$C are shown in Fig.~\ref{fig:carbon_alpha}. The $P_x$ results have a negative sine-like shape, with nearly the same amplitude for $s$-shell knockout from both kinematic settings as well as for $p$-shell knockout at large negative $p_{\rm miss}$ (Setting G); the latter however has a more pronounced asymmetry between the regions of positive and negative $\phi_{pq}$. A possible explanation for this asymmetry is that the accepted events with positive and negative $\phi_{pq}$  have slightly different kinematics from one another, due to acceptance effects. For $p$-shell knockout at small $p_{\rm miss}$ (Setting A) the amplitude is considerably larger. In near-coplanar kinematics   ($\phi_{pq}\approx 0\degree$ or $\pm180\degree$), the measured $P_x$ is consistent with  zero.

The component $P_y$ exhibits a similar $\phi_{pq}$ dependence in $s$-shell knockout (top right panel) and $p$-shell knockout at large
      $|p_{\rm miss}|$ (Setting G, lowest right panel) being approximately constant at
      $-$0.05. However, for $p$-shell knockout at small  
      $p_{\rm miss}$ (Setting A, middle right panel), $P_y$ has a large cosine-like $\phi_{pq}$ dependence.

The theoretical results with (without) the $L\cdot S$ part of the
optical potential are displayed in Fig.~\ref{fig:carbon_alpha} as solid black
(dashed, green online) curves. The calculations predict in contrast to the data a much smaller
$P_x$ for the $s$ shell than shown by the data, moreover with the
opposite sign. For the $p$-shell knockout at Setting G (left lowest panel)
$P_x$ is almost vanishing. For the $p$-shell knockout at Setting
A (left middle panel), $P_x$ shows a distorted sine-like dependence,
similar to the data, but with an about 35\% smaller
amplitude.

For $P_y$ (right panels) the theory agrees much better with the data
than for $P_x$. For $s$-shell knockout there appears
to be an offset by about 0.05 compared to the data. In particular
for $\phi_{pq}$ near $\pm 180\degree$, the theory predicts a positive value
in contrast to the data. On the other hand, for $p$-shell knockout
(middle and lower right panels) the agreement is quite good. 
For the $p$-shell knockout in Setting A, the amplitude of $P_y$ increases as $p_{\rm miss}$
approaches zero, both in the data and in the calculations,
as shown in the supplementary material.

When performing the calculations with the $L\cdot S$ term of the potential switched off (dashed curves, green online), both components of the induced polarization are nearly zero, except for $p$-shell knockout at $p_{\rm miss}$ near zero (Setting A).  In that region, the no-$L\cdot S$ curves are not close to zero and show a
     stronger oscillatory behavior but with a smaller amplitude, about
     one third of the one with the $L\cdot S$ term included.

\section{Conclusions}
\label{sec:conclusions}
We have measured in $A(e,e'\vec p\,)$ the induced polarization components $P_x$ and $P_y$ for both $^{12}$C and $^2$H, greatly extending the kinematic range of previous $P_y$-only measurements for both nuclei.  
Within the regions where our kinematics overlap in $p_{\rm miss}$ and $\phi_{pq}$ with the existing experiments, our data are consistent with the other experiments.  We find that in the regions where the induced polarization depends on $\phi_{pq}$, the dependence is sine-like for $P_x$ and cosine-like for $P_y$.  

For both nuclei, $P_x$ is consistent with zero for near-coplanar events, as expected.  However, the theoretical results for $P_x$ for both nuclei predict a considerably smaller $P_x$ than reflected in the data, moreover in some cases with the opposite sign.  

For $^2$H, the induced polarization is near zero at small $|p_{\rm miss}|$.   $P_x$ shows no significant deviation from zero except at the negative-$p_{\rm miss}$ settings, where it shows a dependency on $\phi_{pq}$.  $P_y$ becomes increasingly negative at larger $|p_{\rm miss}|$.  
This is consistent with the calculations, although the
data show a significantly steeper decrease at negative $p_{\rm miss}$.
 $P_y$ shows no significant dependence on $\phi_{pq}$ at any of the kinematic settings in this work.  In the calculations, most of the deviation of $P_y$ from zero comes from the $L\cdot S$ interaction.  

For $^{12}$C, the measured  $\phi_{pq}$ dependence of $P_x$ shows
a pronounced negative sine-like shape in both $s$- and $p$-shell
knockout which is not reflected in the theoretical results. For the $s$-shell
knockout, $P_y$ has no strong dependence on either $p_{\rm miss}$ nor $\phi_{pq}$.  However, for $p$-shell knockout from $^{12}$C, $P_y$ is strongly dependent on $\phi_{pq}$ at small $p_{\rm miss}$, and $P_x$ is also more strongly dependent on $\phi_{pq}$ than in any other region explored in this work.  
This behavior at small $p_{\rm miss}$ for $p$-shell knockout is reproduced by the calculations, wherein the large dependence on $\phi_{pq}$ is mainly due to the spin-orbit term of the FSI.  

The new data presented in this work provide a rich opportunity to further fine-tune the $L\cdot S$ part of the optical and $NN$ potentials.

\section{Acknowledgements}
We would like to thank the Mainz Microtron operators and technical crew for the excellent operation of the accelerator. This work is supported by the Israel Science Foundation (Grants 390/15, 951/19) of the Israel Academy of Arts and Sciences, by the Israel Ministry of Science, Technology and Spaces, by the PAZY Foundation (Grant 294/18), by the Deutsche Forschungsgemeinschaft (Collaborative Research Center 1044), by the U.S. National Science Foundation (PHY-1205782, PHY-1505615), and by the Croatian Science Foundation Project No. 8570. We acknowledge the financial support from the Slovenian Research Agency (research core funding No.~P1\textendash 0102).  
\FloatBarrier
\section{Supplementary Material}
Supplementary material may be found at \url{insert.url.here.com}.

\section{References}
\bibliographystyle{elsarticle-num}

\addcontentsline{toc}{section}{\refname}\small{\bibliography{induced}}
\clearpage
\newpage
\renewcommand\appendixpagename{Supplementary Materials}
\appendixpage
\begin{appendices}

\setcounter{figure}{0}   
\renewcommand{\thefigure}{S.\arabic{figure}}

\renewcommand{\thesection}{S.\arabic{section}}
%
\section{Determination of false asymmetries using elastic $ep$ measurements}
In order to determine the false asymmetry of the polarimeter, we performed measurements using a hydrogen target, for which the induced polarization is expected to vanish.  The settings for both spectrometers were similar to Setting A, except that we scanned through several proton-momentum settings in order to cover a large range of the focal plane with $ep$ data (for which the proton's momentum and angle are correlated with one another).  The event samples from each of the proton spectrometer momentum settings were combined.  

As noted in the paper, we performed the fit for the false asymmetry by maximizing the log likelihood
\begin{equation}
\log \mathcal{L} = \sum\limits_{\rm events} \log\left[1+\vec A^{\,\rm T} \cdot\left(\begin{array}{c}-\sin{\Phi_{\rm FPP}} \\\cos{\Phi_{\rm FPP}} \\0\end{array}\right)\right],
\end{equation}
for the $ep$ event sample, where $\Phi_{\rm FPP}$ is the azimuthal angle of the secondary scattering and $\vec A$ is  the false asymmetry in the focal plane coordinate system, parameterized as
\begin{equation}
\vec A = \left(\begin{array}{c}a^x_{0} + a^x_{1}\phi_{\rm vth}\\a^y_{0} + a^y_{1}\theta_{\rm vth}  \\0\end{array}\right),
\end{equation}
where $\theta_{\rm vth}$ and $\phi_{\rm vth}$ are the incident angles of the proton trajectory extrapolated from the VDCs to the HDCs.  $a^x_0$, $a^x_1$, $a^y_0$, and $a^y_1$ are the fitted coefficients.

\begin{figure*}[h]
\begin{center}
{\huge $^1{\rm H}(e,e'\vec p\,)$}

\includegraphics[width=\textwidth]{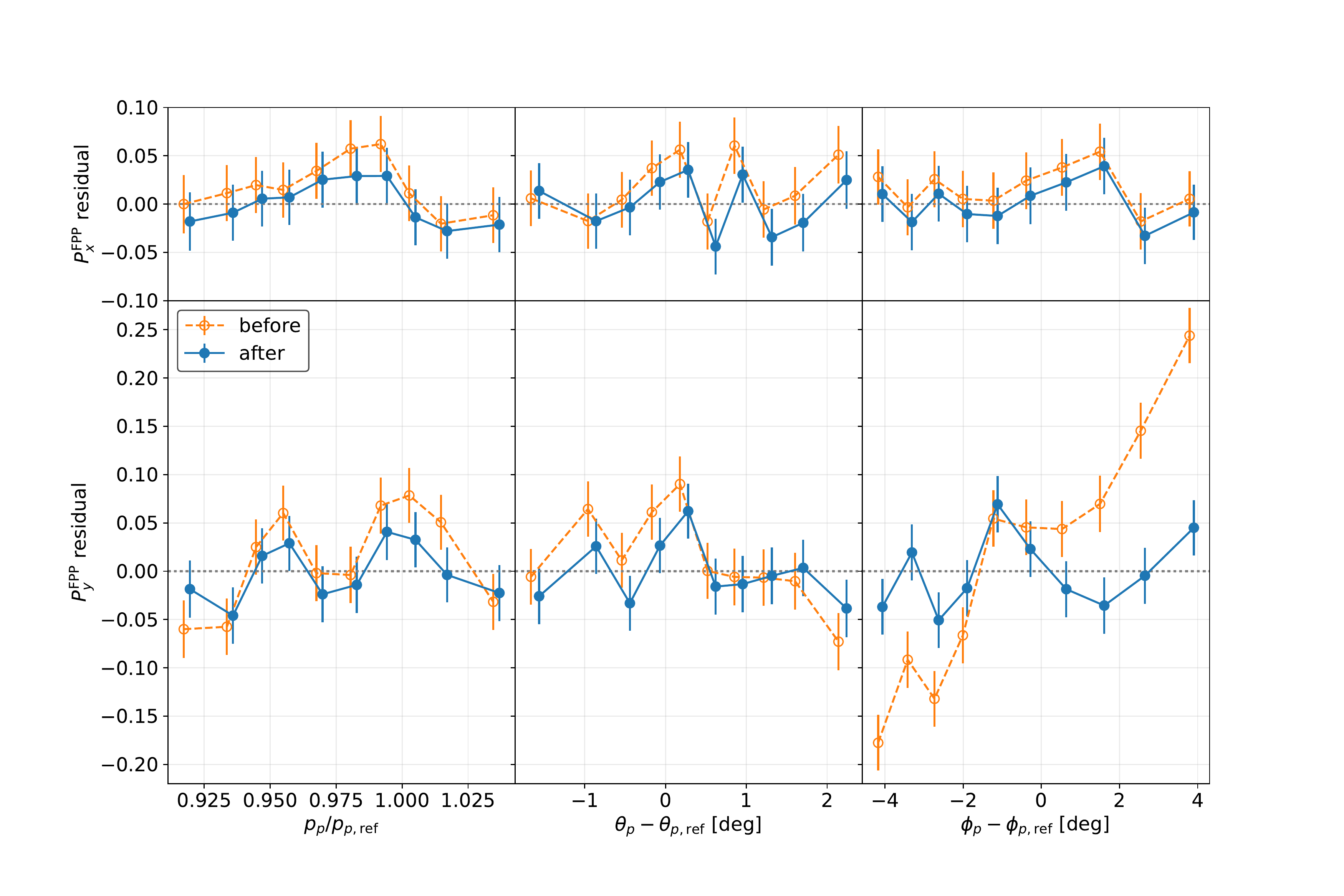}
\end{center}

\caption{The residual induced polarizations in focal-plane coordinates, $P_x^{\rm FPP}$ (top panels) and $P_y^{\rm FPP}$ (bottom panels) measured for $ep$ scattering, without (orange, open circles, connected by a dashed line) and with (blue, filled circles, connected by a solid line) corrections for the false asymmetry.  These are presented in slices of the relative momentum of the proton (left), polar angle (middle), and azimuthal angle (right)}
\label{fig:false_asymmetry}
\end{figure*}

To validate the fits for the false asymmetry, we divided the $ep$ event sample into slices by three of the proton kinematic variables: its momentum (divided by the reference momentum of the spectrometer setting), its polar angle $\theta_p$, and its azimuthal angle $\phi_p$, and performed fits for the components $P_x^{\rm FPP}$ and $P_y^{\rm FPP}$ of the induced polarization in the focal-plane coordinates for each slice.  These fits were performed with (blue filled circles, connected by solid lines) and without (orange open circles, connected by dashed lines) the corrections for the false asymmetry, as shown in Fig.~\ref{fig:false_asymmetry}.  The corrected values of $P_x^{\rm FPP}$ (top) and $P_y^{\rm FPP}$ (bottom)  are consistent with zero.  The  uncorrected values of $P_x^{\rm FPP}$ are slightly higher than zero, whereas those of $P_y^{\rm FPP}$ are large, especially when $|\phi_p-180\degree|$ is large, indicating that the corrections for the former are small while those of the latter are large.  

The $ep$ sample used for the false-asymmetry determination were taken in 2012, just before the $^2$H measurements at Settings A-D were obtained.  We used three checks to validate the long-term stability of the false-asymmetry parameterization between the four run periods.  First, we compared the $^{12}$C data for at Setting G taken in the 2015 run period with the data taken at the same setting during the 2017 run period, and found that these are consistent within error of each other.  Second, the $^2$H data at Settings A and F (which are very similar kinematics to one another), taken in 2012 and 2016 respectively, are consistent with one another within error.  Finally, we checked that the results for $P_x$ (in interaction-point coordinates) at $\phi_{pq}$ near $0\degree$ and $\pm 180\degree$ in all four run periods are consistent with zero within error, as expected.  Together, these checks indicate that the false asymmetry is stable between the run periods.  

\section{Comparison of $^{12}$C data to MIT results}
\begin{figure}[ht]

\begin{center}
{\huge $^{12}{\rm C}(e,e'\vec p\,)$}
\end{center}
\includegraphics[width=\columnwidth]{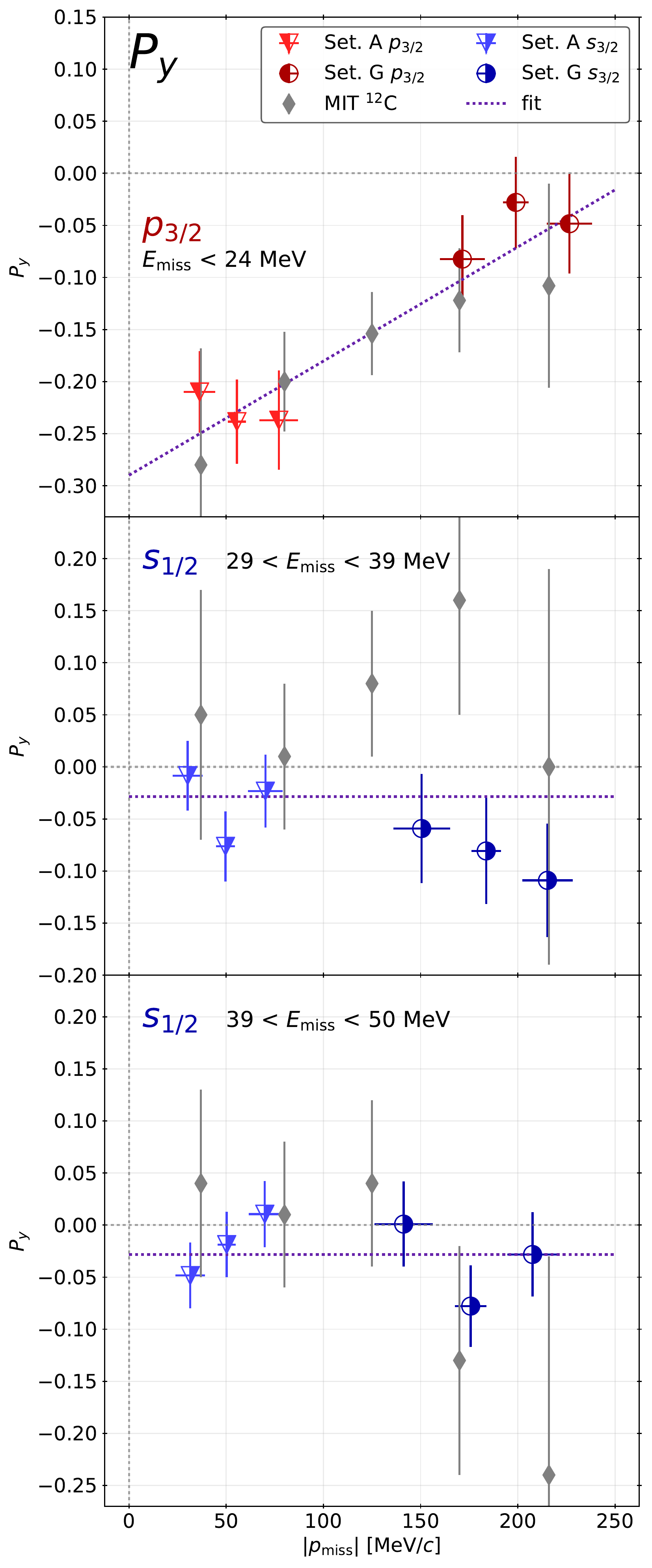}
\caption{
Comparison between the measured $P_y$ for near-coplanar ($\phi_{pq}\approx180\degree$) events in this work and the MIT measurements \cite{Woo}.  The cut used in our data for this plot was $|\phi_{pq}|>135\degree$.  Following \cite{Woo}, the events have been partitioned in slices by $E_{\rm miss}$.  A purple dotted line indicates a combined fit of our data with those of \cite{Woo}.
}
\label{fig:Woo}
\end{figure}

In Fig.~\ref{fig:Woo}, we compare our $^{12}$C measurements (half-filled triangles and circles) to those from MIT \cite{Woo} (filled diamonds) which were taken at $\phi_{pq}$ centered at 180$\degree$.  For this comparison, we selected only the events with $|\phi_{pq}|>135\degree$ for consistency with the the kinematics of the MIT measurements. Following \cite{Woo}, we plot our $s$-shell data separately for two bins in $E_{\rm miss}$: 29 to 39 MeV/$c$ and 39 to 50 MeV/$c$.  The data are shown in bins of $|p_{\rm miss}|$.  

For the $p$-shell knockout, both our high-$|\phi_{pq}|$ data and those of \cite{Woo} have large negative $P_y$ at small $|p_{\rm miss}|$ and small negative $P_y$ at large $|p_{\rm miss}|$.  We fitted our data and those of \cite{Woo} to a straight line and show this fit as a dashed green line in the top panel of Fig.~\ref{fig:Woo}.  Both datasets are consistent with the fit within error ($\chi^2=3.3$ with 9 degrees of freedom; $p_{\rm val}=0.95$).  

For the $s$-shell knockout, we took the weighted average (horizontal-line fit) of $P_y$ for our measurements and those of \cite{Woo} and obtained $P_y^{\rm avg} = -0.03$.  This is shown as a purple dashed line in the middle and bottom panels of Fig.~\ref{fig:Woo}.  Both our data and those of \cite{Woo} are consistent within uncertainty of this value ($\chi^2=19.7$, with 21 d.o.f.; $p_{\rm val}=0.54$).

\section{$\phi_{pq}$-dependence of the induced polarization in $^2$H}
\begin{figure*}[h]
\begin{center}
{\huge $^2{\rm H}(e,e'\vec p\,)$}

\includegraphics[width=\columnwidth]{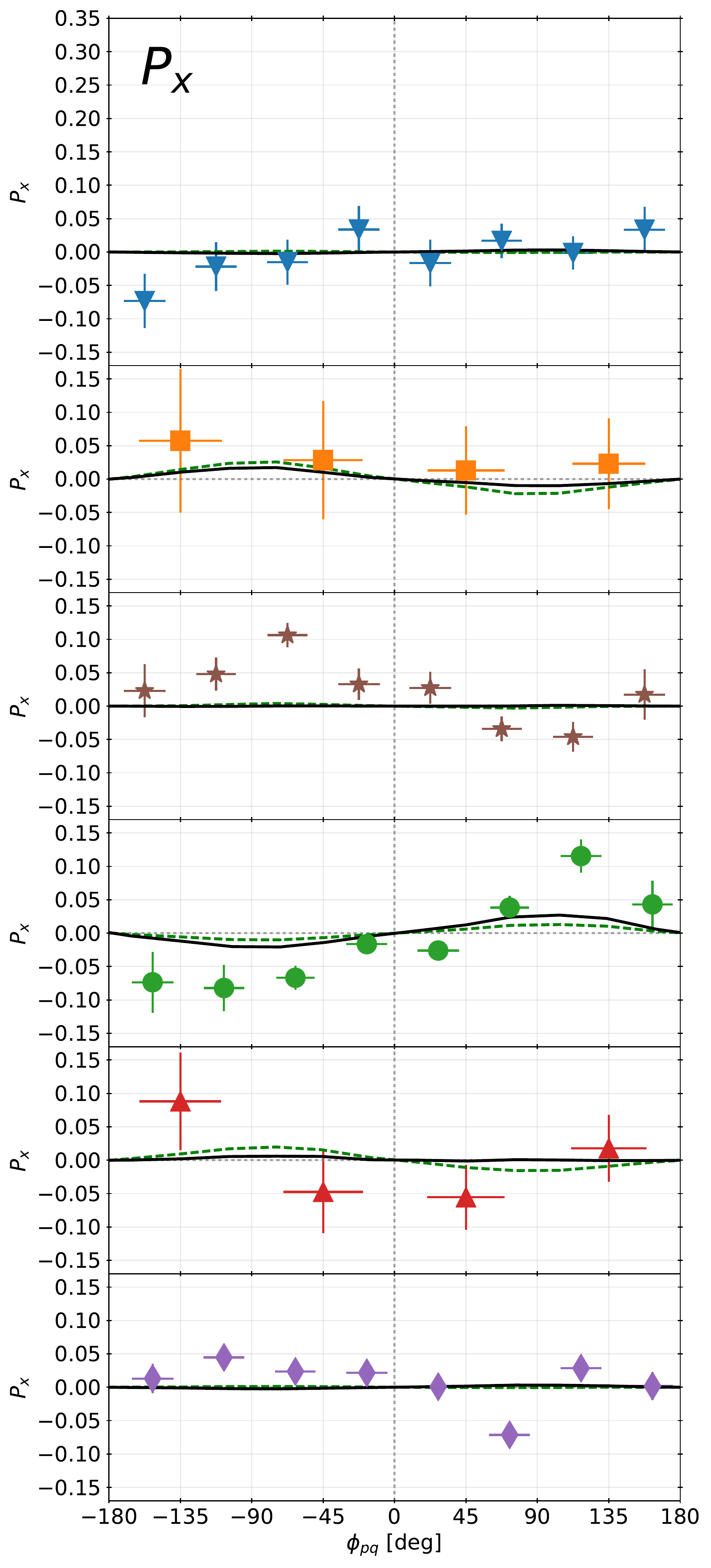}
\hspace{6mm}
\includegraphics[width=\columnwidth]{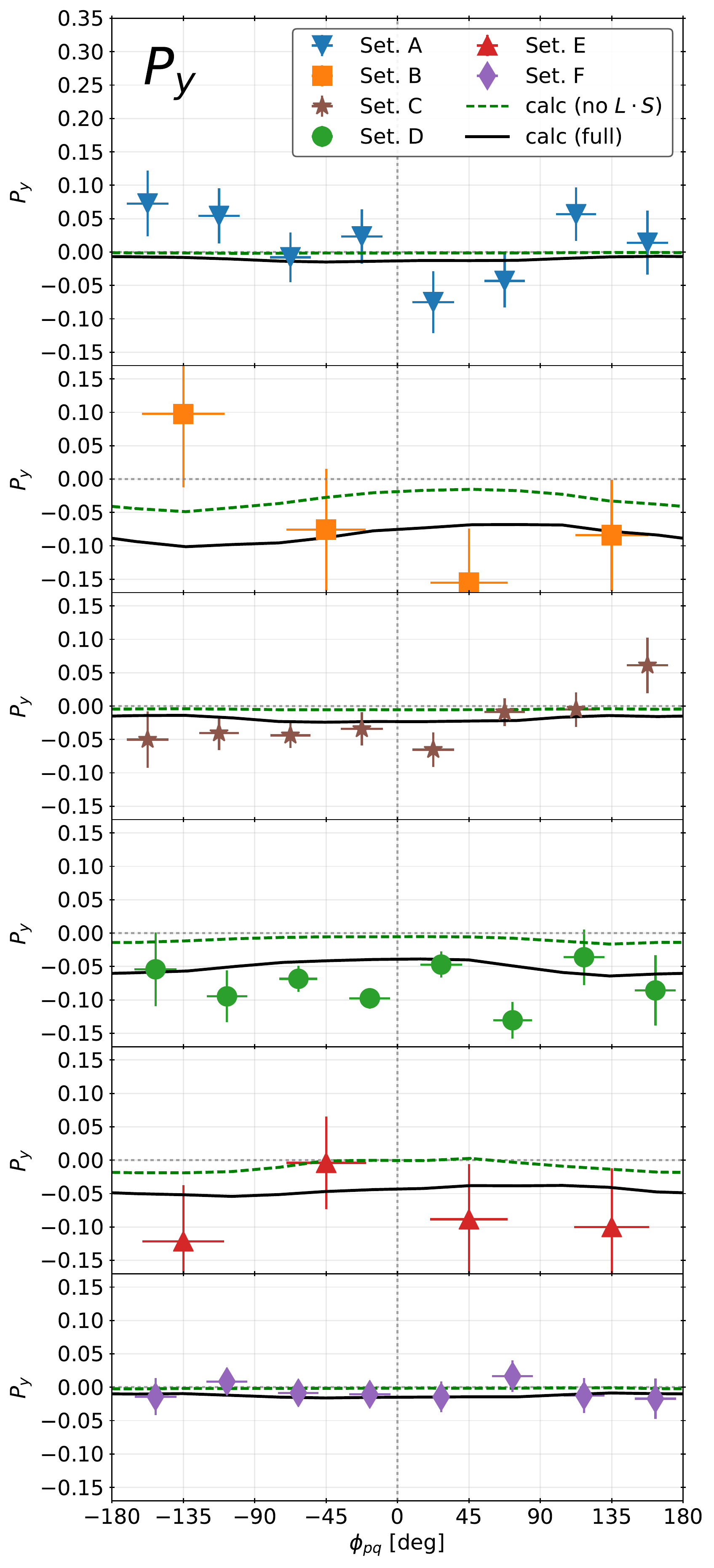}
\end{center}

\caption{The measured induced polarizations, $P_x$ (left panels) and $P_y$ (right panels) for $^{2}$H, plotted versus $\phi_{pq}$.  The data are shown on separate panels for each kinematic setting.  The calculations with (without) the $L\cdot S$  interaction are shown as  solid black (green dashed) curves. 
}
\label{fig:deuteron_alpha}
\end{figure*}

In Fig.~\ref{fig:deuteron_alpha}, we show the dependence of $P_x$ and $P_y$ on $\phi_{pq}$ for all six kinematic settings in both the data and the calculations. For $P_x$, there is no statistically significant variation in the data except in Settings C and D, as noted in the paper.  $P_y$ shows no significant dependence on $\phi_{pq}$, while the calculation-curves are nearly flat for all six settings except Setting D (large negative $p_{\rm miss}$, at $Q^2=0.18$ (GeV\!/\!$c$)$^2$).  

%

\section{$\phi_{pq}$-dependence of the induced polarization in $^{12}$C at small $p_{\rm miss}$}
As noted in the paper, the data at Setting A for $p$-shell knockout has a large $\phi_{pq}$-dependence for $P_y$, while the rest of the data show no such dependence.  This was observed in both the data and the calculations.  We also found that $P_x$ for this subset of the data is also larger in magnitude than for the rest of the data.  

\begin{figure*}[h!]
\begin{center}
{\huge $^{12}{\rm C}(e,e'\vec p\,)$}

\includegraphics[width=\columnwidth]{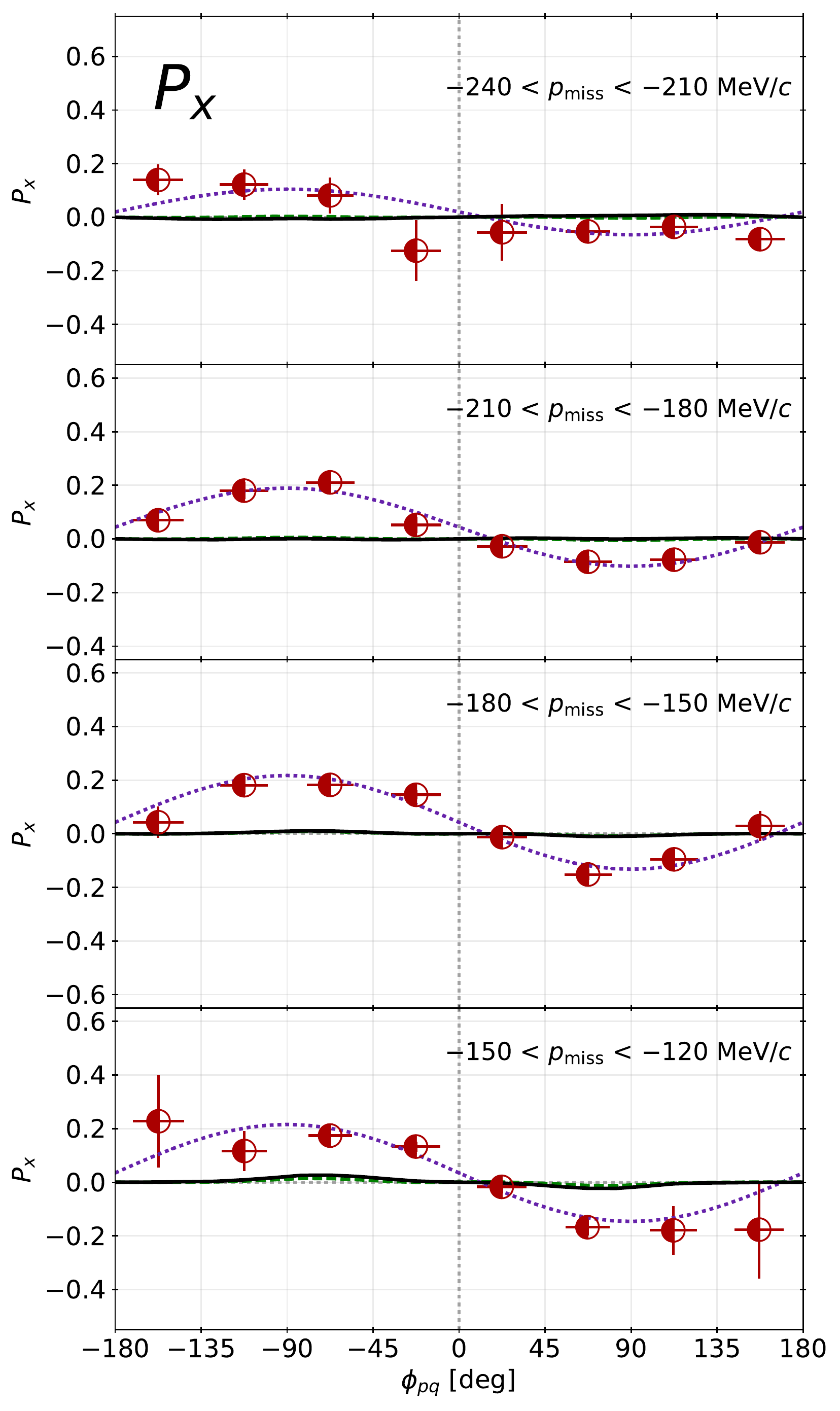}\hspace{6mm}
\includegraphics[width=\columnwidth]{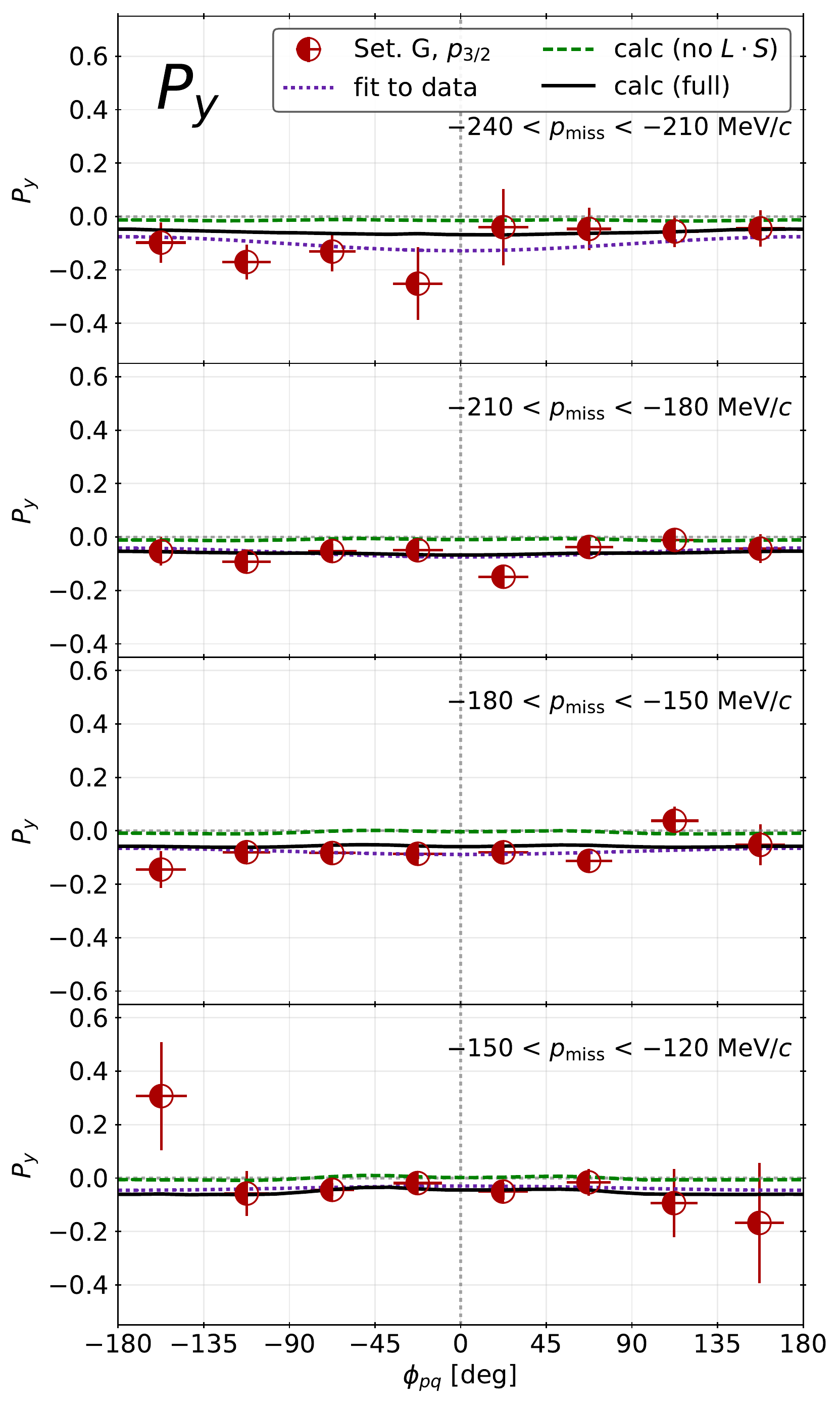}
\end{center}
\caption{$P_x$ (left panels) and $P_y$ (right panels) as functions of $\phi_{pq}$, in several slices of $p_{\rm miss}$ for the $p$-shell knockout in $^{12}$C at Setting A.~  The calculations with (without) the $L\cdot S$  part of the optical potential are shown as  solid black (green dashed) curves.  Fits of the data to $P_x = a_x+b_x\sin(\phi_{pq})$ and $P_y = a_y+b_y\cos(\phi_{pq})$ are shown for each slice (purple dotted curves).
}
\label{fig:carbon_alpha_slices_G}
\end{figure*}

\begin{figure*}[h!]
\begin{center}
{\huge $^{12}{\rm C}(e,e'\vec p\,)$}

\includegraphics[width=\columnwidth]{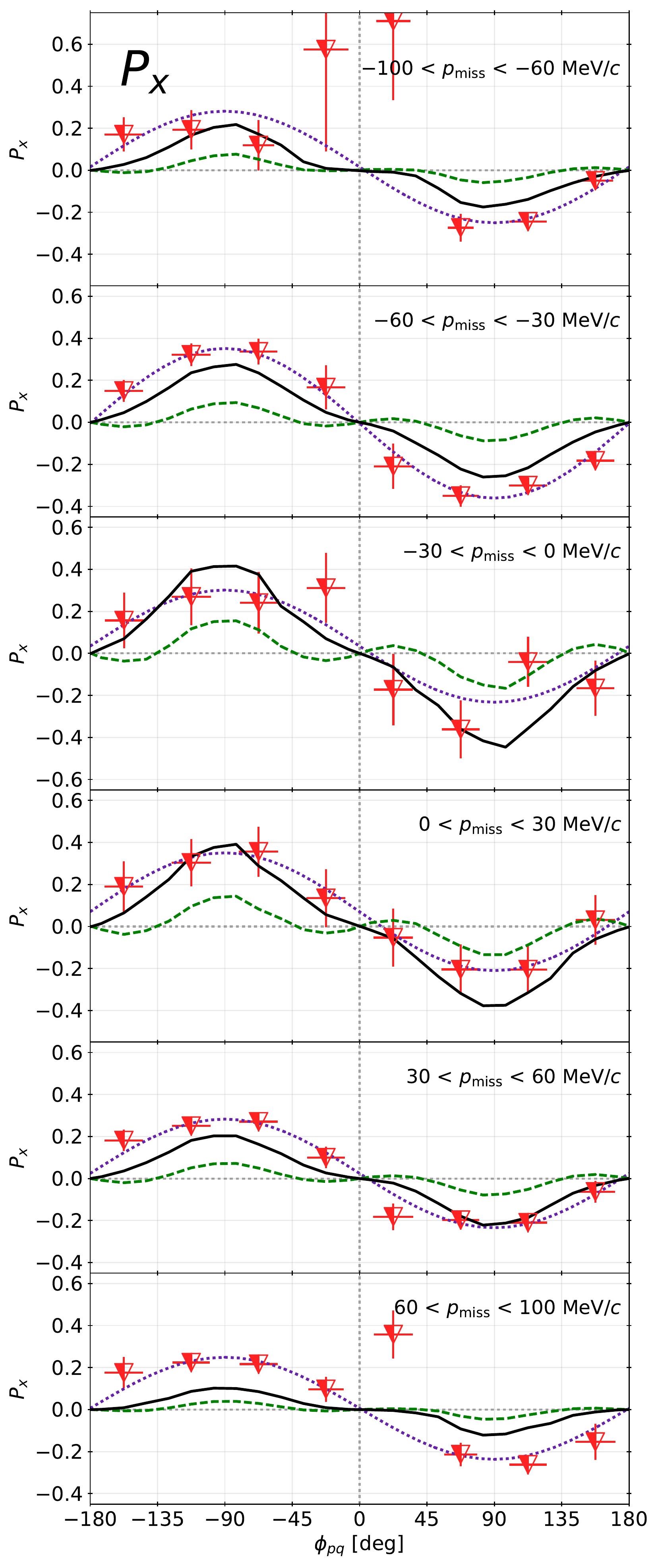}\hspace{7mm}
\includegraphics[width=\columnwidth]{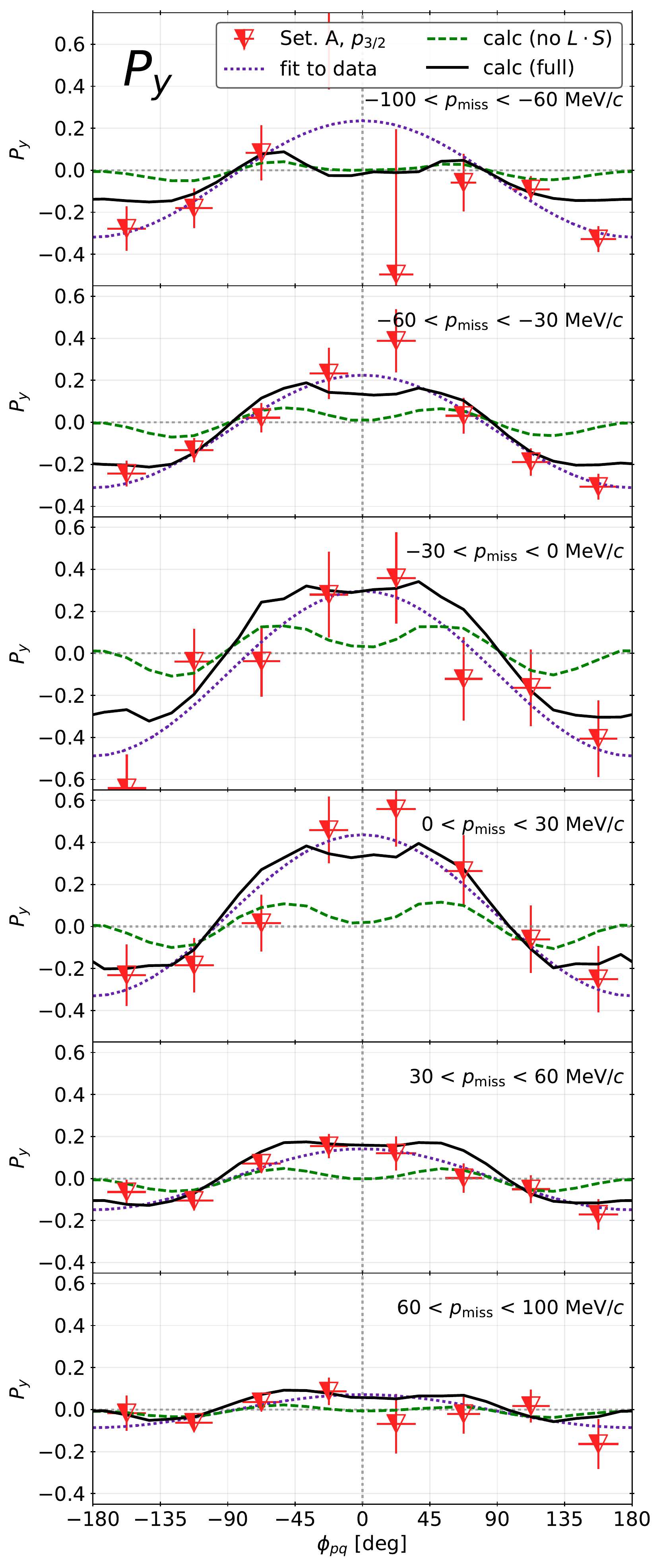}
\end{center}
\caption{Same as Fig.~\ref{fig:carbon_alpha_slices_G}, for Setting A.  
}
\label{fig:carbon_alpha_slices}
\end{figure*}

In order to determine if this $\phi_{pq}$ dependence is correlated with $p_{\rm miss}$, we partitioned the $p$-shell data into slices by $p_{\rm miss}$, and plot the $\phi_{pq}$ dependence for each slice in Figs.~\ref{fig:carbon_alpha_slices_G} (Setting G) and \ref{fig:carbon_alpha_slices} (Setting A).  A purple dotted curve represents a fit to the data in each slice by the equations
\begin{equation}
P_x = a_x + b_x\sin\phi_{pq}
\end{equation}
and   
\begin{equation}
P_y= a_y + b_y\cos\phi_{pq}.
\end{equation}

\begin{figure}[h]

\begin{center}
{\huge $^{12}{\rm C}(e,e'\vec p\,)$}
\end{center}
\includegraphics[width=\columnwidth]{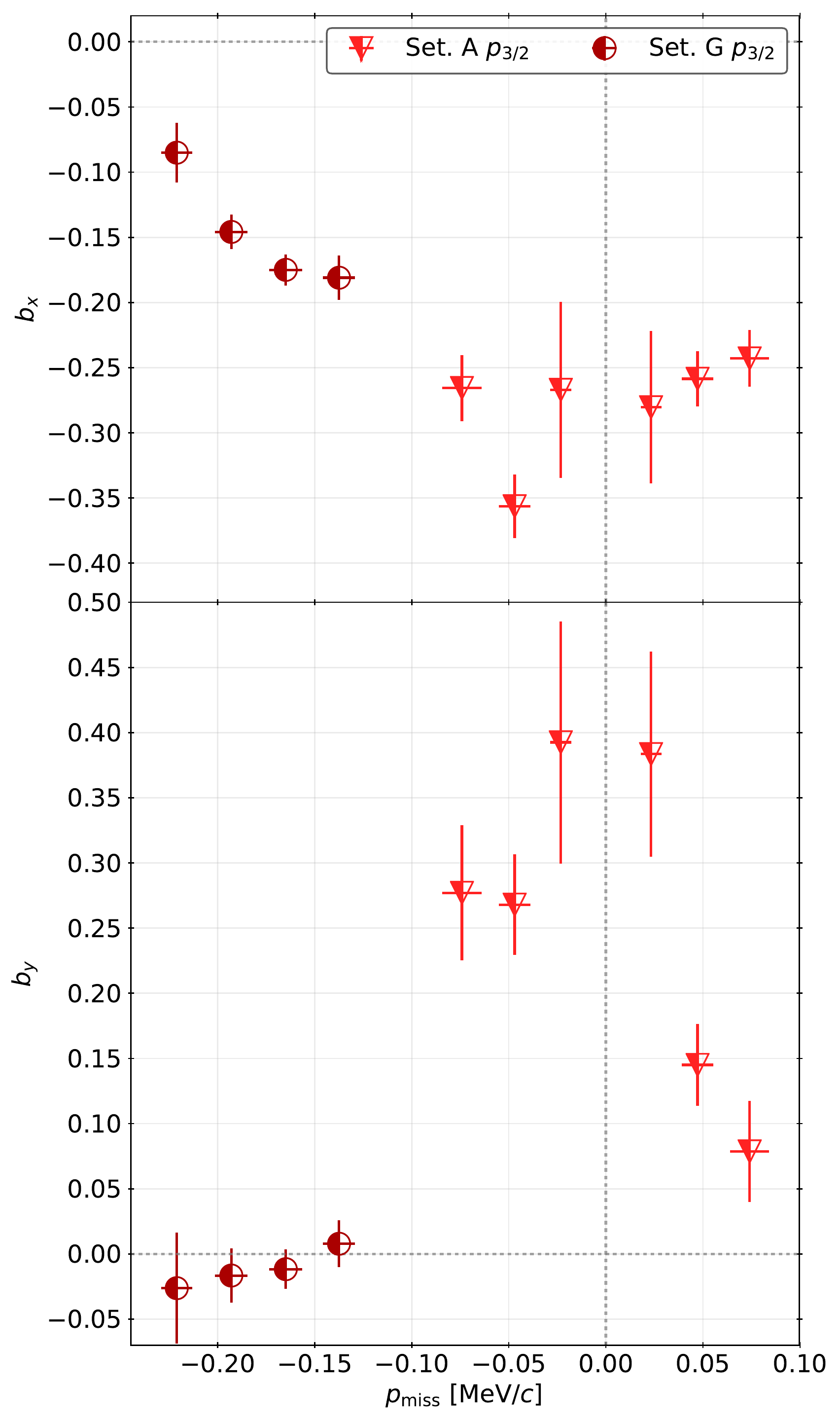}
\caption{
The parameters $b_x$ and $b_y$ from the fits $P_x = a_x + b_x\sin\phi_{pq}$ and $P_y = a_y + b_y\sin\phi_{pq}$.  These parameters describe the $\phi_{pq}$ dependence of $P_x$ and $P_y$ respectively.  
}
\label{fig:bxby}
\end{figure}

The values of $b_x$ and $b_y$ (which consider the $\phi_{pq}$ dependence of $P_x$ and $P_y$) are shown in Fig.~\ref{fig:bxby}.  The fit parameter $b_x$ is small at large negative $p_{\rm miss}$ (Setting G), and increases in absolute value to a negative plateau as $p_{\rm miss}$ approaches 0 in Setting A. The fit parameter $b_y$ is near zero for large $p_{\rm miss}$ (Setting G) and reaches a sharp peak at $p_{\rm miss}=0$ (middle of Setting A).  The calculations curves for $P_x$ at Setting G are flat with respect to $\phi_{pq}$, whereas the data suggest otherwise, as reflected by the non-zero $b_x$ in the fits.


\end{appendices}
\end{document}